\documentclass{aa}  

\usepackage{graphicx}
\usepackage{natbib}
\bibpunct{(}{)}{;}{a}{}{,} % to follow the A&A style
\usepackage[varg]{txfonts}
\usepackage{stfloats}

\def\kms{$\rm km\, s^{-1}$}

\def\Ha{H$\alpha$~}
\def\fe2{[Fe\,\textsc{ii}]}
\def\teste{[N\,\textsc{ii}]}
\def\o3{O\,[\textsc{iii}]}
\def\h2{H$_{\rm 2}$}

\def\cm3{$\rm cm^{-3}$}

\def\s3{[S\textsc{iii}]}

\def\Ha{H$\alpha$}

\begin{document} 

\title{Black hole feeding and star formation in NGC\,1808}
 %\subtitle{}
  \author{A. Audibert \inst{1,2,3,4}  \and F. Combes\inst{1,5} \and S. Garc{\'i}a-Burillo\inst{6} \and L. Hunt \inst{7} \and A. Eckart \inst{8}  \and S. Aalto \inst{9} \and V. Casasola \inst{10} \and  F. Boone \inst{11} \and M. Krips \inst{12} \and S. Viti \inst{13,14}  \and S. Muller \inst{9} \and K. Dasyra \inst{15}  \and P. van der Werf \inst{13} \and S. Mart{\'i}n\inst{16,17}}

   \institute{Observatoire de Paris, LERMA, PSL University, Sorbonne University, CNRS, Paris, France\\
        \email{anelise.audibert@iac.es}
          \and
          National Observatory of Athens (NOA), Institute for Astronomy, Astrophysics, Space Applications and Remote Sensing (IAASARS), GR–15236, Greece  
          \and Instituto de Astrof{\'i}sica de Canarias, Calle V{\'i}a L{\'a}ctea, s/n, E-38205, La Laguna, Tenerife, Spain 
          \and Departamento de Astrof{\'i}sica, Universidad de La Laguna, E-38206, La Laguna, Tenerife, Spain
          \and
                        Coll{\`e}ge de France, 11 Pl. Marcelin Berthelot, 75231, Paris  
          \and
             Observatorio Astron{\'o}mico Nacional (OAN-IGN)-Observatorio de Madrid, Alfonso XII, 3, 28014 Madrid, Spain
                \and
                         INAF - Osservatorio Astrofisico di Arcetri, Largo E. Fermi, 5, 50125, Firenze, Italy \and I. Physikalisches Institut, Universit{\"a}t zu K{\"o}ln, Z{\"u}lpicher Str. 77, 50937, K{\"o}ln, Germany
\and Department of Space, Earth and Environment, Chalmers University of Technology, Onsala Space Observatory, SE-43992 Onsala, Sweden
\and INAF - Istituto di Radioastronomia, via Piero Gobetti 101, 40129, Bologna, Italy
\and CNRS, IRAP, 9 Av. colonel Roche, BP 44346, 31028, Toulouse Cedex 4, France
\and IRAM, 300 rue de la Piscine, Domaine Universitaire, F-38406 Saint Martin d’H{\`e}res, France
\and Leiden Observatory, Leiden University, PO Box 9513, 2300 RA Leiden, Netherlands
\and Dep. of Physics and Astronomy, UCL, Gower Place, London WC1E 6BT, UK
\and Dep. of Astrophysics, Astronomy \& Mechanics, Faculty of Physics, National and Kapodistrian University of Athens, Panepistimiopolis Zografou, 15784, Greece, and National Observatory of Athens, Institute for Astronomy, Astrophysics, Space
Applications and Remote Sensing, Penteli, 15236, Athens, Greece
\and European Southern Observatory, Alonso de C{\'o}rdova, 3107, Vitacura, Santiago 763-0355, Chile
\and Joint ALMA Observatory, Alonso de C{\'o}rdova, 3107, Vitacura, Santiago 763-0355, Chile  \\ }          

   \date{Received November, 2020; accepted 2021}

% \abstract{}{}{}{}{} 
% 5 {} token are mandatory
 
\abstract{We report on Atacama Large Millimeter Array (ALMA) observations of CO(3-2) emission in the Seyfert2/starburst galaxy NGC~1808, at a spatial resolution of 4\,pc. Our aim is to investigate the morphology and dynamics of the gas inside the central 0.5\,kpc and to probe the nuclear feeding and feedback phenomena. We discovered a nuclear spiral of radius 1 '' = 45\,pc. Within it, we found a decoupled circumnuclear disk or molecular torus of a radius of 0.13'' = 6\,pc. The HCN(4-3) and HCO$\rm^+$(4-3) and CS(7-6) dense gas line tracers were simultaneously mapped and detected in the nuclear spiral and they present the same misalignment in the molecular torus. At the nucleus, the HCN/HCO$^+$ and HCN/CS ratios indicate the presence of an active galactic nucleus (AGN). The molecular gas shows regular rotation, within a radius of 400\,pc, except for the misaligned disk inside the nuclear spiral arms. The computations of the torques exerted on the gas by the barred stellar potential reveal that the gas within a radius of 100\,pc is feeding the nucleus on a timescale of five rotations or on an average timescale of $\sim$60\,Myr. Some non-circular motions are observed towards the center, corresponding to the nuclear spiral arms. We cannot rule out that small extra kinematic perturbations could be interpreted as a weak outflow attributed to AGN feedback. The molecular outflow detected at $\geqslant$250\,pc in the NE direction is likely due to supernovae feedback and it is connected to the kpc-scale superwind.} 


   \maketitle
%
%________________________________________________________________

\section{Introduction}

%__________________________________________________________________

The principal issue in understanding how supermassive black holes (SMBHs) are growing at the center of spiral galaxies is the ability to unveil the processes feeding the nuclear activity and -- when gas is present in the circumnuclear environment -- the transferring of the angular momentum outward to allow the gas to inflow. These processes are complex and take place on multiple scales: Gravity torques from non-axisymmetric features in the stellar potential, such as bars or spirals, can play a major role at 0.1-1\,kpc scales \citep [e.g.,][]{buta96}, but the gas may be stalled at Lindblad resonances. The prolongation of the angular momentum transfer at smaller radii requires a cascade of embedded structures, such nuclear bars \citep [e.g.,][]{Shlosman1989,hop10}. Such detailed processes require us to determine the morphology and kinematics of the gas at a high spatial resolution at a 10\,pc scale. The feeding processes are regulated by feedback from the active gactic nucleus (AGN) and they are able to entrain significant molecular outflows, which have been frequently observed over the past decade \citep [e.g.,][]{fer10,cicone14,flu19,Veilleux2020}. However, feedback might also come from star formation at 0.1-1kpc scales. Thus, to disentangle the AGN origin of the feedback, a high spatial resolution is also required \citep{aalto16,santi19}. 

At 10\,pc scales, given that there are shorter dynamical timescales involved than in the main disks due to the chaotic and often turbulent accretion episodes into the SMBH, the gas might change its orientation and the presence of decoupled molecular tori (with the inner disks kinematically decoupled) is ubiquitous \citep{santi1068,santigatos, Alonso2018,ah19,combes19}. This decoupling makes the feedback more efficient, given that the AGN wind or jet is able to entrain molecular gas from the disk.

The physical mechanisms needed to feed galaxy nuclei at the 10\,pc scale are not yet very well known, but the advent of ALMA has recently opened new windows for tracing the molecular gas at the required resolution and sensitivity \citep{santi1068,santi16tor,santi19, combes19, ah19}. Bars and nuclear bars were already proposed long ago to act as dynamical mechanisms \citep{Shlosman1989} and confirmed at kpc down to 100\,pc scale \citep{combes01,santi12}; recently, nuclear bars acting on gaseous nuclear spirals have also been observed to be very efficient at the 10\,pc scale \citep{combes1566,ane613}. The goal is now to find more cases to confirm this mechanism. The high spatial resolution is also key to discovering AGN feedback and
distinguishing it from supernovae (SN) feedback \citep{cicone14,fer15,salak16}. In many galaxies, both occur simultaneously
\citep{rupke05,veilleux13,kal14}, but winds and molecular outflows starting at radii below the 10\,pc scale would point 
towards AGN feedback.

\begin{figure}
 \resizebox{\hsize}{!}{\includegraphics{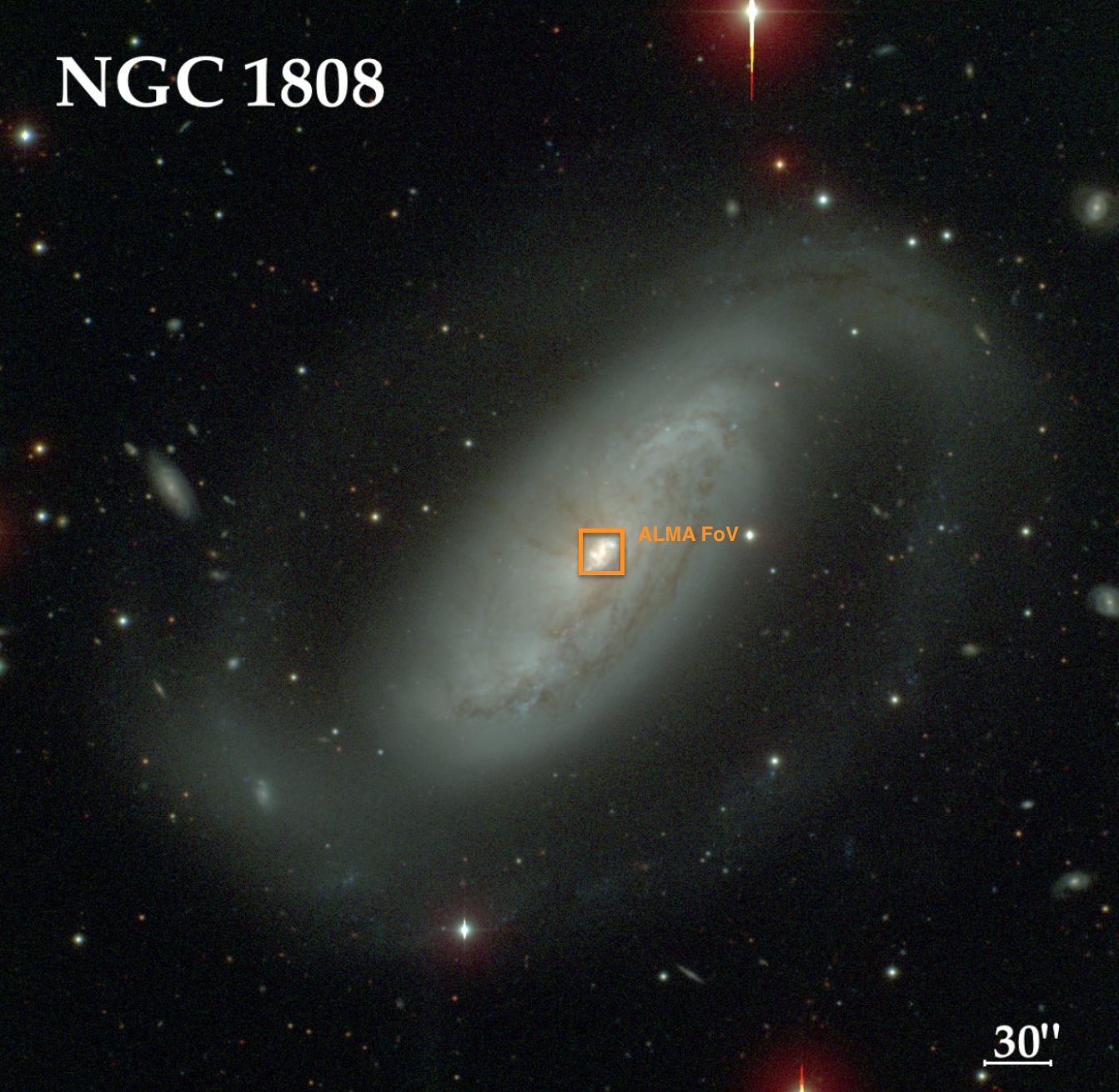}}
\caption{Large-scale (8.9'$\times$8.9') optical image of NGC\,1808 obtained with the 2.5-meter Ir{\'e}n{\'e}e du Pont telescope at Las Campanas Observatory. The ALMA 18\arcsec FoV is shown in the orange square for comparison. Credits: Carnegie-Irvine Galaxy Survey \citep{ho11}.}
\label{cgs}
\end{figure}

In order to investigate the physical phenomena of AGN feeding and feedback in this paper, we study the kinematics of the molecular gas in the nearby Starburst/Seyfert 2 galaxy NGC\,1808. This galaxy is remarkable for its very dusty appearance and especially for its well-known system of dark radial filaments; it has also dust lanes  that are likely perpendicular to the disk, parallel to the minor axis, up to 3\,kpc from the major axis. It is known for its peculiar nuclear region of “hot spots” \citep{sp65}. NGC\,1808 is a Starburst/Sy\,2 galaxy located at 9.3\,Mpc (1\arcsec=45\,pc) and it is classified as an SAB(s)a. According to \citet{reif82}, it has an inclination of 57$^\circ$ and a PA=311$^\circ$ in the optical \citep{Dahlem1990}. In Figure~\ref{cgs}, a large-scale (8.9'$\times$8.9') optical image of the galaxy obtained with the Carnegie-Irvine Galaxy Survey (CGS) is shown \citep{ho11}. Atomic gas traced with H\,{\sc i} is concentrated in the galactic bar, disk, and a warped outer disk, indicating a tidal interaction in the past with the neighbour galaxy NGC\,1792, located at a projected distance of $\sim$150\,kpc \citep{kor96}. The prominent polar dust lanes correspond to gas outflow emerging from the nucleus, as revealed in optical studies that indicate an outflow of neutral and ionized gas, as seen in \Ha, \teste, Na\,\textsc{i}\,D and H\,\textsc{i} by \citet{kor93, phi93}. In particular, Na\,\textsc{i}\,D is seen blueshifted in absorption and redshifted in emission; also, $B$-$R$ images show that the dust plumes associated with the outflow reach $\sim$3\,kpc above the plane. 
NGC\,1808 has been classified as a superwind galaxy, similar to M\,82, due to the evidence of large-scale outflows, likely due to the starburst \citep{Dahlem1990, Dahlem1994}.
Near-infrared (NIR) integral-field spectroscopy with the Spectrograph for INtegral Field Observations in the Near Infrared (SINFONI) by \citet{bus17} showed a large gas reservoir and a disturbed gas velocity field that shows signs of inflowing streaming motion in the central $\sim$100\,pc. The main properties of NGC\,1808 are summarized in Table~\ref{prop}.

In this paper, we present ALMA observations in the CO(3-2) line of NGC\,1808, with a spatial resolution up to 4\,pc. Its proximity (9.3~Mpc) and moderate inclination of 57$^\circ$ make NGC~1808 an ideal target to test and refine scenarios of AGN feeding and feedback, and to potentially discover new phenomena controlling gas structures and dynamics within 400 pc, where the dynamical time-scale is smaller than $\sim$ 10 Myr (for $\rm V_{\rm rot}$=200 km/s). Observations are detailed in Section 2 and the results are presented in Section 3. The interpretation in term of torques is discussed in Section 4 and our conclusions are drawn in Section 5.

\begin{table}
\caption{Properties of NGC\,1808}              % title of Table
\centering                                      
\begin{tabular}{l c c}         
\hline\hline   
\noalign{\smallskip}                    
 Parameter &Value & Reference \\  
\hline    
\noalign{\smallskip}                            
$\alpha$J2000$^a$ & 05h07m42.34s   & (1) \\
$\delta$J2000$^a$ &  -37d30m46.98s &  (1) \\
V$\rm _{\rm sys}$ & 995\,km$\rm s^{-1}$ &  (1) \\
RC3 Type & SAB(s)a & (2) \\
Nuclear Activity  & Sy\,2/ Starburst & (1,3) \\
Inclination  & 57$^\circ$ &  (4) \\
Position Angle  & 311$^\circ$,314$^\circ$ & (5,8) \\
Distance$^b$ & 9.3$\pm$2.4\,Mpc & (1) \\
SFR$^c$ & 4.7$\rm\,M_\odot yr^{-1}$ & (1) \\
$\rm M_{\rm H\textsc{i}}$ & 3.4$\rm \times10^{9}\,M_\odot$ & (4) \\
$\rm M_*$  & 2.1$\rm \times10^{10}\,M_\odot$ & (6)  \\
$\rm M_{\rm H2}$ & 2$\rm \times10^{9}\,M_\odot$ & (5)  \\
$\rm L_{\rm IR}$ & 5.1$\times$10$\rm ^{10}L_\odot$ & (7) \\
$\alpha_{\rm cont}$J2000$^d$ & 05h07m42.33s  &  (8) \\
$\delta_{\rm cont}$J2000$^d$ & -37d30m45.88s &  (8) \\
$\alpha_{\rm spiral}$J2000$^e$ & 05h07m42.34s & (8) \\
$\delta_{\rm spiral}$J2000$^e$ & -37d30m45.71s & (8) \\
\hline        
\label{prop}
\end{tabular}
\\
\tablefoot{References:  1: NASA/IPAC Extragalactic Database (NED); 2: \citep{vau91}; 3:  \citet{veron86}; 4: \citet{reif82}; 5:  \citep{Dahlem1990}; 6: \citet{combes19}; 7: \citep{san03}; 8: this work.\\
\tablefoottext{$^a$}{($\alpha_{J2000}$, $\delta_{J2000}$) is the phase tracking center of our interferometric observations.} \\
\tablefoottext{$^b$}{Distance is the median values of z-independent distances from NED \citet{steer17} } \\
\tablefoottext{$^c$}{SFR is derived from infrared luminosity (NED)} \\
\tablefoottext{$^d$}{The RA-DEC positions are the new adopted center, derived from the central continuum peak in this work, with an uncertainty of $\sim$0.1\arcsec\ (see Sec.~\ref{cont_emission})} \\
\tablefoottext{$^e$}{The RA-DEC positions of the nuclear spiral, used as the kinematical center in Sections~\ref{kin} and \ref{torques}.} 
}
\end{table}

\section{Observations}

\begin{table*}[h]
\caption{Observations summary.}
\label{tab:obs}
\begin{center} \begin{tabular}{cccccccc}
   \hline
Date  & $N_{\rm ant}$  & $B_{\rm min}$ / $B_{\rm max}$ $^{(b)}$ & $\theta_{\rm res}$ $^{(c)}$ & $LRAS$ $^{(d)}$ & $PWV$ $^{(e)}$ & $t_{\rm on}$ $^{(f)}$ \\
     & $^{(a)}$      & (m / km)                           & ($\arcsec$)   & ($\arcsec$)  & (mm) & (min)   \\
   \hline 

   21 Apr 2016 $^{(g)}$ & 42 & 15 / 0.6   & 0.36 & 3.9  &  0.8 & 4 \\
   12 Aug 2016 $^{(g)}$ & 38 & 15 / 1.5  & 0.14 & 2.7  &  0.4 & 20 \\

   26 Nov 2016 $^{(h)}$ & 42  & 15 / 0.7  & 0.31 & 2.7  &  0.4 & 11 \\
   03 Aug 2017 $^{(h)}$ & 44  & 21 / 3.6  & 0.07 & 0.8  &  0.6 & 35 \\

\hline
\end{tabular} \end{center}
\mbox{\,} \vskip -.25cm
Notes: $(a)$ Number of 12\,m-antennas in the array;
$(b)$ Respectively minimum and maximum projected baseline;
$(c)$ Synthesized beam with this configuration alone;
$(d)$ Largest recoverable angular scale;
$(e)$ Amount of precipitable water vapor in the atmosphere;
$(f)$ On-source time;
$(g)$ ALMA project number 2015.1.00404.S; $(h)$ ALMA project number 2016.1.00296.S.
\end{table*}

NGC1808 was observed with the ALMA in several array configurations, covering the CO(3-2), CS(7-6), HCN(4-3), and HCO$\rm ^{+}$ lines in band 7 with a final combined resolution of 0.27\arcsec($\sim$12\,pc). We present the summary of the observations below and in Table~\ref{tab:obs}.

First, in ALMA cycle 3 (project ID: \#2015.1.00404.S, PI F. Combes), NGC\,1808 was observed  simultaneously in CO(3-2), HCO$\rm^{+}$(4-3), HCN(4-3) and CS(7-6) for both the compact (TC, baselines 15 to 612m) and the extended (TE, baselines 15 to 1462~m) configurations. The TC configuration was observed in April 2016 with 42 antennas and an on-source integration time of four minutes, providing a synthesized beam of $\sim$0\farcs36. The TE configuration was observed in August 2016 with 38 antennas, on-source integration of 20 minutes and a synthesized beam of $\sim$0\farcs14. The correlator setup, designed to simultaneously observe the four lines, provided a velocity range of 1600\,km/s for each line, but did not center the HCO$\rm^{+}$(4-3) and HCN(4-3) lines (200\,km/s on one side and 1400\,km/s on the other, which is adequate for a nearly face-on galaxy), and 1800\,MHz bandwidth in the continuum.

Then, in ALMA Cycle 4, the observations (project ID:\#2016.1.00296.S, PI F. Combes) were performed in November 2016 and August 2017 at
a higher spatial resolution (0\farcs08 corresponding to $\sim$4\,pc), with the aim to resolve the molecular torus. The tuning configuration of Band 7 was in the CO(3-2), HCO$\rm^{+}$(4-3) and continuum, providing a 
broader spectral window centered at the emission line. The correlator setup was selected to center the CO(3-2) and the
HCO$\rm^{+}$ lines in the 2\,GHz bandwidth. The compact configuration (TM2, baselines 15 to 704~m) was observed with 42 antennas for an integration time of 11 minutes and a synthesized beam of 0\farcs31 and the extended (TM1, baselines 21 to 3638m), with 44 antennas for 35 minutes on-source and a synthesized beam of 0\farcs07.

The observations were centred on the nucleus, with a single pointing covering a field of view (FoV) of 18\arcsec. The galaxy was observed in dual polarization mode with 1.875 GHz total bandwidth per spectral window, and a channel spacing of 0.488 MHz corresponding to $\sim$0.8\,km/s, after Hanning smoothing. The flux calibration was done with radio quasars (J0522-3627, J0519-4546, J0453-3949) close to the position in the sky of the target, which are regularly monitored at ALMA, and resulted in about 10\% accuracy.

The data from Cycle 3 and 4 were calibrated and concatenated with the \textsc{CASA} software (version from 4.5.3 to 4.7.2, \citep{casa}, and the imaging and cleaning were performed with the \textsc{GILDAS} software \citep{gildas}. The analysis were made in \textsc{GILDAS} together with \textsc{python} packages \citep[radio-astro-tools, APLpy, PySpecKit][]{radiotools, aplpy,pyspec}. The spectral line maps were obtained after subtraction of the continuum in the $uv$-plane using the tasks \textsc{uv\_continuum} and \textsc{uv\_subtract}. Differently than in the paper presenting the data by \citet{combes19} -- where we used only the most extended configurations
(TM1+TE) -- in this work, we combine all the configurations. In the case of the emission line datacubes, 
the \textsc{CLEAN}ing was performed using the Hogbom method and a natural weighting, excluding the TM1 configuration (due to very high noise outside the nuclear region) in order to achieve the best sensitivity, resulting in a synthesized beam of 0\farcs30$\times$0\farcs24 for the concatenated data cube.  The data cubes were produced with a resolution of 10.16\,km/s (11\,MHz).
For the continuum maps, we selected the line emission free channels and concatenated all the configurations (TE+TC+TM1+TM2), and we provide two images according to resolution: the \textit{contA} map, using natural weighting, yielding a synthesized beam of 0\farcs2 and \textit{contB} using uniform weighting and achieving a better angular resolution, with a synthesized beam size of 0\farcs08 ($\sim$4\,pc).
The total integration time provided an rms of 57$\mu$Jy/beam in the continuum and in the line channel maps, a 0.35\,mJy/beam per channel of 10\,km/s. The final maps were corrected for primary beam attenuation. Very little CO(3-2) emission was detected outside the  full-width half-power (FWHP) primary beam ($\sim$17\arcsec).

\section{Results}

\subsection{Continuum emission}\label{cont_emission}

\begin{figure}
\centering
 \resizebox{\hsize}{!}{\includegraphics{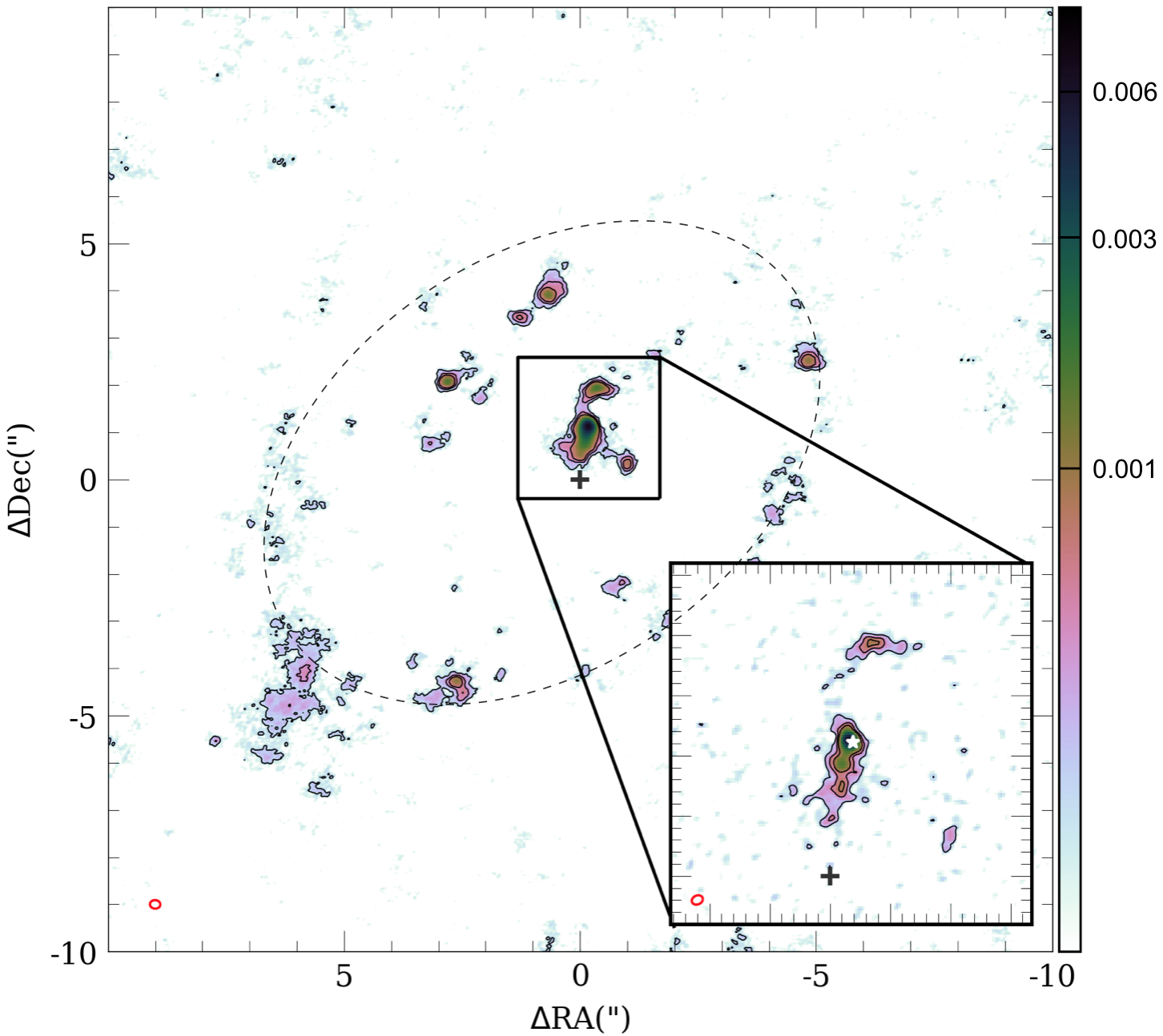}}
\caption[Continuum emission at 0.87\,mm.]{Continuum emission at 0.87\,mm. The central 20\arcsec$\rm \times$20\arcsec\ is shown in the main panel for the natural weighting image (\textit{contA}). The red ellipse at the bottom left corner indicates the synthesized beam size of 0.22\arcsec$\times$0.18\arcsec at PA=86$^\circ$. The contours are 3, 5, and 9 times the $\sigma_{\rm rms}$. The zoom-in of the central 3\arcsec$\rm \times$3\arcsec\ at high resolution (0.09\arcsec$\times$0.07\arcsec at PA=115$^\circ$, \textit{contB}) is shown in the  bottom-right corner. The black cross indicates the adopted phase center, while the white star represents the peak of the continuum emission, assumed to be the AGN position (see Table~\ref{prop}). The color scale is in Jy/beam.}
\label{cont}
\end{figure}

\begin{figure*}
\centering
\includegraphics[width=17cm]{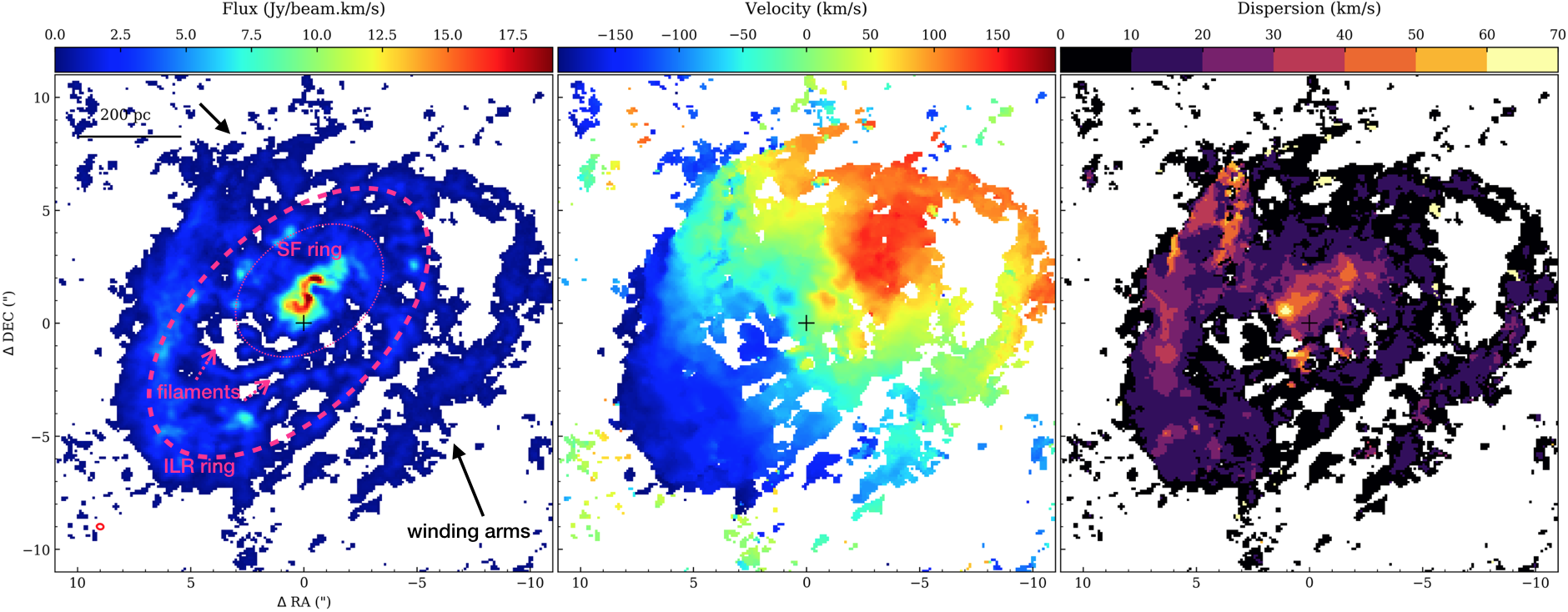}
\caption{CO(3-2) moments maps of NGC\,1808 in the central 22\arcsec\ ($\sim$1\,kpc). \textit{Left:} Integrated intensity map (0$th$ moment) of CO(3-2). \textit{Middle:} Mean velocity (1st moment). \textit{Right:} Velocity dispersion (2nd moment). The synthesized beam is indicated as the red ellipse in the bottom left corner of the integrated intensity map. The main morphological features of the CO(3-2) emission are indicated in the 0th moment map: the ILR ring (dashed pink ellipse) projected to the galaxy inclination, $i$=57$^\circ$; the SF ring (dotted pink ellipse) projected to the same $i$=52$^\circ$ of the ionized and H$\rm_2$ gas detected with SINFONI by \cite{bus17}; the pink arrows indicate the position of the filamentary structure connecting the rings; and the black arrows show the outer winding spiral arms. The black cross corresponds to the phase center of our observations, listed in Table~\ref{prop}. 
}
\label{moms}
\end{figure*}

Radio continuum emission in NGC\,1808 was reported by \citet{col94} using 3.6\,cm ($\sim$8 GHz) radio observations with the Very Large Array (VLA). These authors found a  family of compact radio sources along the ring in the central region of the galaxy and the nuclear component has a flux of 62.8\,mJy. Compared to the VLA measurements at  6\,cm of \citet{saikia90}, the spectral indices of the compact sources in the ring are consistent with almost all of them being supernovae remnants (SNRs) or H\,{\sc ii} regions. 
In the nuclear core, ALMA band 3 observations at $\sim$93\,GHz and 2.6\arcsec resolution found a spectral index $\alpha_{\rm nuc}\sim-1$ \citep{salak16, salak17}, typical of synchrotron emission. They suggested that the nuclear region is dominated by a high-energy source, namely, supernovae explosions or a low-luminosity AGN (LLAGN). The results are consistent with the findings of \citet{Dahlem1990}, who measured a spectral index of $\alpha_{\rm nuc}$=-0.9 from 20 and 6~cm radio images.

Our 0.87\,mm (350 GHz) continuum observation in band 7 at 0.2\arcsec resolution (\textit{contA}) is shown in Figure \ref{cont}. We find a nuclear component and some compact regions along the ring. In the zoomed panel at the bottom right, we show the continuum at higher resolution (4\,pc, \textit{contB}). The white star in the figure indicates the peak of the continuum emission, adopted as the AGN position, listed in Table~\ref{prop}. The continuum is a point source inside the nuclear spiral and may also correspond to a torus, of radius  0.13\arcsec =6\,pc \citep{combes19}.

 The frequency around 100 GHz is the minimum intensity between the synchrotron emission with negative slope and the dust emission with positive slope. It is not possible to compare all intensities at all frequencies because of different resolutions and apertures. However, we can compare the 3 fluxes within 2\arcsec\ diameter at 8, 350 and 500 GHz. On the synchrotron side, the 8 and 93 GHz can be compared, and the slope is -1 in this region.

In the nuclear region, for the 3.6\,cm (8.3GHz) VLA observations, the integrated flux is 39.2~mJy in a 2\arcsec\ diameter aperture \citep{koti96}. At 350~GHz, the higher resolution continuum map, \textit{contB}, shows a maximum in flux of $S_{\rm peak}(contB)$=3.51~mJy/beam and the 0.22\arcsec$\times$0.18\arcsec\ continuum map, \textit{contA}, has $S_{\rm peak}(contA)$=6.78~mJy/beam. We smoothed our 350~GHz continuum map at the same resolution and the integrated flux in a 2\arcsec\ diameter region is 27.8~mJy.
 The continuum at 500~GHz observed with ALMA in Band 8 at 0.825\arcsec$\times$0.59\arcsec~resolution is presented by \citet{salak19}. At the AGN position, the continuum peaks at 54~mJy and smoothing our 350~GHz continuum at the same resolution of the 500~GHz map, we derive a spectral index of $\alpha$=3.6.
  In summary, all fluxes are compatible with the sum of two power-laws, a synchrotron of slope -1, and a dust emission of slope +3.6, leaving little space in the middle for a free-free emission, within the error bars.

\subsection{Molecular gas distribution and morphology}

In the appendix, we present the channel maps for the CO(3-2) observations. Figure~\ref{chans} displays 40 of the CO(3-2) channel maps, with a velocity range of 400 km/s and a velocity resolution of 10.2\,km/s.  The channels show evidence of a regular velocity field in a patchy ring at a radius $\sim$7\arcsec (315\,pc), that is most prominent in the south part and another broken ring at $\sim$3.5\arcsec(160\,pc). They are connected by multiple spiral arms or filaments. Inside the star-forming (SF) ring at 3.5\arcsec, that is, also detected in the ionized and H$\rm_2$ gas by \citet{bus17}, at the very center, a 2-arm spiral structure is clearly seen, which is also observed in CS, HCN and HCO$^+$, as we further discuss in Section~\ref{dense}.

The moment maps of the CO(3-2) line (clipping the emission at $<$3$\rm\sigma_{\rm rms}$) are shown in Figure~\ref{moms}.  The integrated intensity (zero-moment) map in the left panel of Figure~\ref{moms} shows that the CO emission follows the $\sim$300\,pc SF circumnuclear ring. \citet{salak16} and \citet{salak17} have mapped the CO(1-0) and CO(3-2) emission with ALMA at resolution of 1\farcs17$\times$0\farcs77 and 1\farcs04$\times$0\farcs56, respectively. They also detected molecular spiral arms and a 500 pc pseudo-ring, however, they noted a structure that they called ``molecular torus,''  which is a double peak structure at radius 30\,pc and a circumnuclear disk in the central 100 pc. At our higher spatial resolution, this double peak interpreted as a torus is actually the two-armed nuclear spiral evidently seen in the central 100\,pc. The evidence of inflowing gas through this trailing spirals will be discussed in Section~\ref{inflow}. We also find a torus in a smaller compact structure, of radius 6\,pc, inside the nuclear spiral \citep{combes19}.

\begin{figure}
 \resizebox{\hsize}{!}{\includegraphics{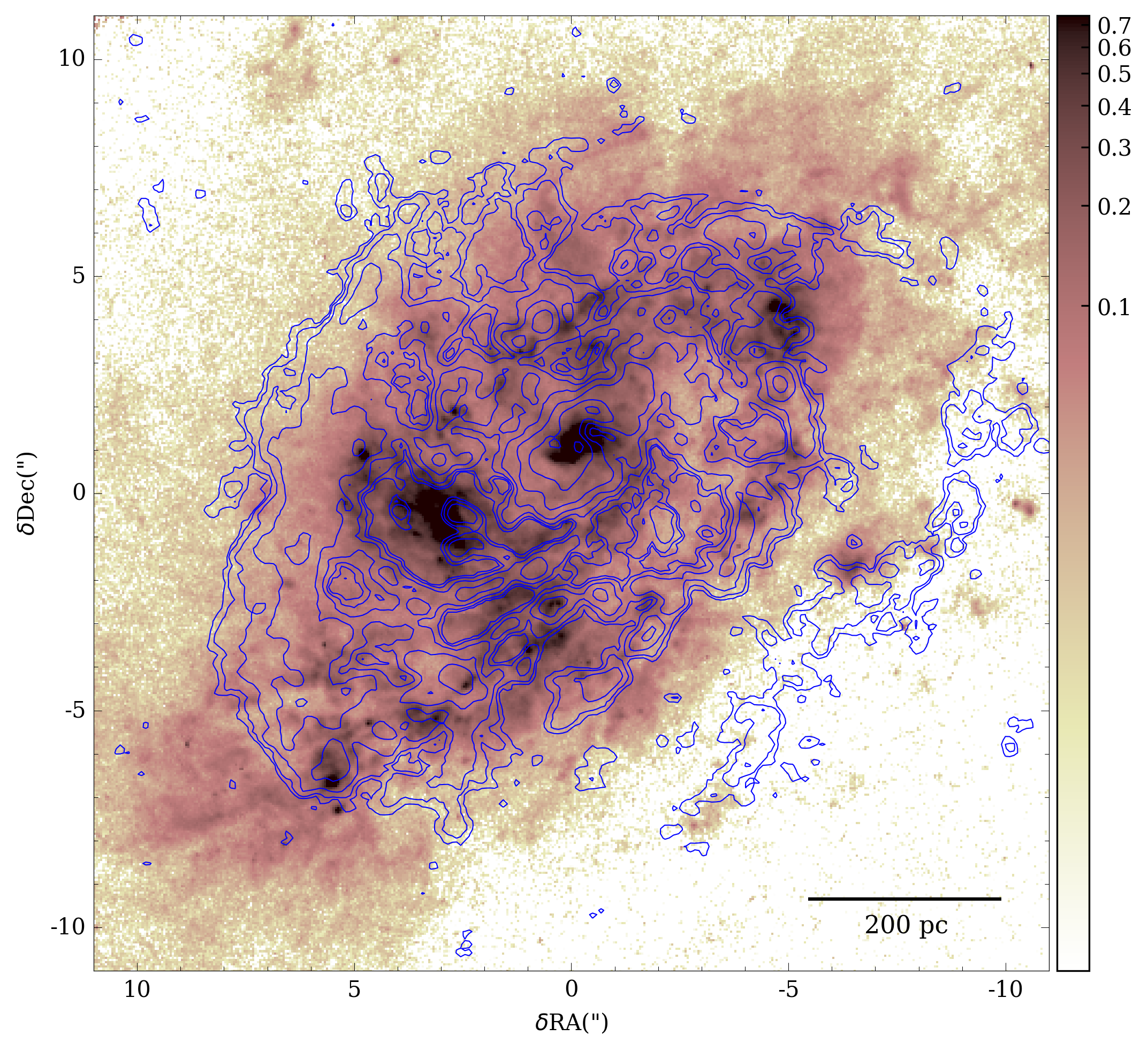}}
\caption[HST image of NGC\,1808 with CO(3-2) contours.]{WFC/F658N HST image and the CO(3-2) contours overlaid.}
\label{hst}
\end{figure}

Figure~\ref{hst} shows the CO(3-2) contours superposed onto the Hubble Space Telescope (HST) maps in the F658N filter\footnote{The HST image was aligned to the ALMA astrometry, the peak emission in the HST image was re-centered to the AGN position in Table~\ref{prop}. The resolution is 0.05\,\arcsec\/pixel.}. The CO and optical maps show a remarkable similarity in morphology; the molecular ring seen in the CO emission coincides with the dusty nuclear ring in the HST image and the winding arms are the beginning of the characteristic dust lanes along the bar. 

In the middle panel of Figure~\ref{moms}, the intensity-weighted velocity (first-moment) map shows a clear rotation pattern in the galaxy plane, with velocities peaking between $-200<v<150$\,\kms from the systemic velocity ($v_{\rm sys}$=995\,\kms). 
In the central region, at radius r$\lesssim$1\arcsec , there is a clear change in PA, pointing to a circumnuclear disk that is  kinematically decoupled from the main disk, in which the position angle is being tilted from 314 to close to 270$^\circ$.
Close to the AGN position, the velocity dispersion is about ($\sigma\sim$40\,km/s), as displayed in the right panel of Figure~\ref{moms} (second-moment map). Along the nuclear spiral the velocity dispersion ranges from 20$\sim$66\,km/s and the average dispersion along the ring is $\sim$10\,km/s. Moreover, we can distinguish a disturbance in the southest part of the disk, with an increased dispersion $\gtrsim$35\,km/s and in the NE direction reaching up to $\sigma{\rm _v}\approx$60\,km/s.

\subsection{CO luminosity and H$\rm_2$ mass}\label{miss}

In order to estimate the missing flux, we compare the ALMA observations to the central spectrum obtained with the 15-m single dish obtained with the Swedish-ESO Submillimeter Telescope (SEST) in CO(1-0) and CO(2-1) over a 43\arcsec~ and 22\arcsec~ FoV, respectively. Towards the central position, \citet{aalto94} found a CO(2-1) spectrum peaking at T$_{\rm mb}$= 0.84\,K with a FWHM=268\,km/s, yielding a total integrated flux of 227\,K\,km/s (4751\,Jy\,km/s), in a beam of 22\arcsec. Their beam is very similar to our FoV of 18\arcsec, therefore, we can compare the fluxes since both FoV's encompasses the entire central disk.

We used a brightness temperature ratio of $\rm r_{\rm 32}=T_{\rm 3-2}/T_{\rm 2-1}$ of 0.66$\pm$0.02 and $\rm r_{\rm 21}=T_{\rm 2-1}/T_{\rm 1-0}=1.12\pm$0.01, derived from the CO(2-1)/CO(1-0) ratio of the SEST observations in the galaxy center \citep{aalto94}. This is expected for thermalized excitation and a dense molecular medium. In that case, the CO(3-2) flux should be higher than the CO(2-1), as we could expect the flux density $\rm S_{\rm \nu}\propto\nu^2$ in the Rayleigh-Jeans approximation, for gas at a temperature higher than 25~K and density higher than 10$^4$ cm$^{-3}$:
\begin{equation}
S_{\rm \nu} \rm{(Jy)}=\frac{2\kappa T_{\rm B}}{c^2} \nu^2 {\Omega}^2 = \frac{2.65}{\lambda^2 \rm{(cm^2)} } T_{\rm B} \rm{(K)} {\Theta_{\rm b}}^2\rm(arcmin),
\end{equation}
where $\kappa$ is the Boltzmann constant, $T_{\rm B}$ is the brightness temperature, $c$ is the speed of light, $\Omega$ is the beam solid angle. For a Gaussian primary beam with a FWHP beam size $\Theta_{\rm b}$, the beam solid angle is $\Omega=\pi\Theta_{\rm b}/4ln2$ and provides the right hand expression in term of the wavelength $\lambda$ in centimeters, the beam size $\Theta_{\rm b}$ in arcminutes and $T_{\rm B}$ in Kelvins. Using these values, the expected CO(3-2) intensity is $\sim$4723\,Jy\,km/s in a 18\arcsec\ beam. When integrated over the spectral range ($v\sim\pm$200\,km/s), the integrated emission in our ALMA FoV of 18\arcsec\ is 2594\,Jy\,km/s. Therefore, we should expect some missing flux by a factor up to $\sim$40-50\%, taking into account the uncertainties of the $r_{\rm 32}$ and $r_{\rm 21}$ ratios \citep[deconvolution uncertainties were not taken into account in][]{aalto94}. 

The mass of molecular can be estimated from the integrated flux $\rm S_{\rm CO}\Delta V$ using the equation from \citet{sol05}:
\begin{equation}
{L_{\rm CO}}^\prime(\rm{K\,km\,s^{-1}\,pc^2}) = 3.25\times10^7 \frac{S_{\rm CO}\Delta V}{1+z}  \left( \frac{D_{\rm L}}{\nu_{\rm rest}} \right)^2 
,\end{equation}
where $\nu_{\rm rest}$ is the rest frequency of the observed line, in GHz, $D_{\rm L}$ is the luminosity distance, in Mpc, $z$ is the redshift and $\rm S_{\rm CO}\Delta V$ is the integrated flux in Jy\,km/s. The molecular mass then can be calculated using the CO-to-H$_2$ conversion factor $M(H_{\rm 2})=\alpha_{\rm CO} r_{\rm 13} {L_{\rm CO}}^\prime$. 

\begin{figure*}
\centering
\includegraphics[width=17cm]{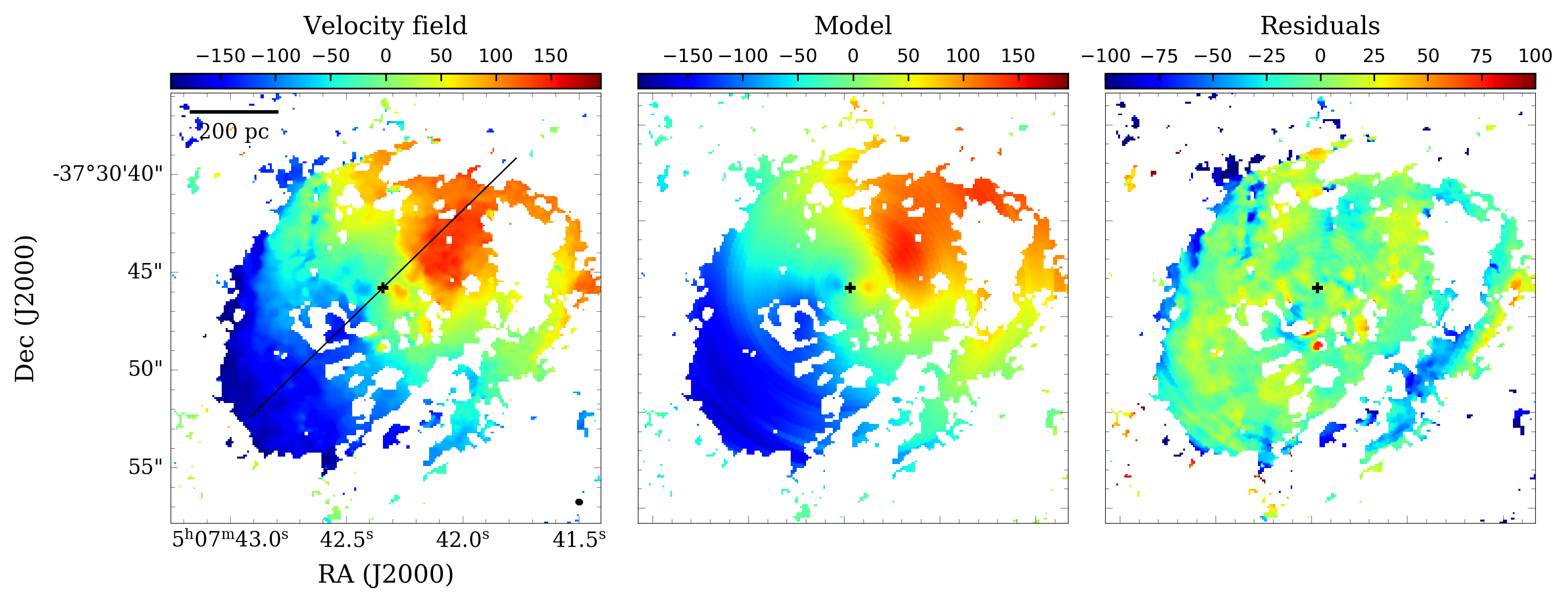}
\caption{ Model of the velocity field using the tilted rings approach. {\it Left:} Velocity map of NGC\,1808 clipped at $>$3$\sigma$. In the {\it middle} panel the best fit using the tilted-ring model \citet{rog74} and the residuals, after subtracting the model from the data, are shown on the {\it right}. The straight line in the left panel represents the kinematical PA=314$^\circ$ of the major axis derived from the best fit and the black cross indicates the center of the nuclear spiral adopted as the center position in the rotation curve fitting.}
\label{tilted}
\end{figure*}

We find a CO luminosity of ${L_{\rm CO}}^\prime=$6.1$\times10^7$\,K\,km/spc$^2$ and molecular mass of 2.7$\times$10$^8$M$_\odot$ in our FoV, assuming a Milky Way-like conversion factor for CO-to-H$_2$  of $\alpha_{\rm CO}$=4.36\,M$_{\rm \odot}(\rm K\,km\,s^{-1}\,pc^2)^{-1}$ \citep[e.g.,][]{bol13} and the CO(3-2) being thermalized and optically thick (r$_{\rm 13}$=1). If instead we use r$_{\rm 13}$=1.35 given the r$_{\rm 32}$ and r$_{\rm 21}$ ratios reported by \citet{aalto94}, we obtain a molecular mass of 3.6$\times$10$^8$M$_\odot$. In comparison, using the SEST CO(1-0) observations with 43\arcsec\ beam resolution integrated over an mapped area of $\Delta\Omega\approx3.4\times10^{-7}$sr$\approx$4\,$arcmin^2$ and standard conversion factor, \citet{Dahlem1990} derived a total molecular mass of $\rm M_{\rm H2}\approx$1.5$\times$10$^9$\,M$_\odot$  (Table~\ref{prop}), adapting to our adopted distance. For the values of the SEST CO(2-1) spectrum in the central 22\arcsec\ beam reported by \citet{aalto94}, it gives a mass of 9.75$\times$10$^8$\,M$_\odot$ for a CO(2-1)/CO(1-0) ratio of $1.12$. The larger mapped area of the CO(1-0) observations from \citet{Dahlem1990} provides the total molecular mass, while the $\rm M_{\rm H2}$ estimates from \citet{aalto94} over the 22\arcsec beam indicate the nuclear molecular content. Our rather smaller value derived with the ALMA high resolution observations is due to lack of single-dish observations to detect weak extended emission (missing flux of $\sim$40-50\%).

\subsection{CO(3-2) kinematics}\label{kin}

To derive the CO(3-2) kinematics, we have assumed a simple model for the rotation curve, proposed by \citet{ber91}, which considers that the gas is on circular orbits in a plane, $v_{\rm c}=Ar/(r^2 +c^2)^{p/2}$, where for $p=1$ the velocity curve is asymptotically flat and $p=3/2$ the system has a finite total mass, therefore, we expect $1\leq p \leq 3/2$. The observed radial velocity at a position $\rm (R,\Psi)$ on the plane of the sky can be described as:

\begin{eqnarray}\label{eq:ber}
v(R,\Psi)= v_{\rm sys} +~~~~~~~~~~~~~~~~~~~~~~~~~~~~~~~~~~~~~~~~~~~~~~~~~~~~~~~~~~~~~~~~~~~~ \\
\frac{A R cos(\Psi -\Psi_0)sin(\theta)cos^p(\theta)}{\left \{ R^2 \left[ sin^2(\Psi -\Psi_0)+ cos^2(\theta)cos^2(\Psi -\Psi_0) \right] +c^2 cos^2(\theta) \right \} ^{p/2} } \nonumber,
\end{eqnarray}

\noindent where $\theta$ is the inclination of the disk (with $\theta$ = 0 for a face-on disk), $\Psi_0$ is the position angle of the line of nodes, $v_{\rm sys}$ is the systemic velocity, and $R$ is the radius. In addition, $A$, $c$, and $p$ are parameters of the model. 

From the velocity map, the dominant velocity feature appears to be attributed to the circular rotation in the disk. We used the tilted-ring model \citep{rog74}, which consists of dividing the velocity field in concentric rings in radii $\Delta r$ and each ring is allowed to have an arbitrary $v_c$, $i,$ and PA. For each radius, we can independently fit the parameters of Equation~\ref{eq:ber} to the observed velocity field. We show the results of the tilted-ring fitting to the velocity map in Figure~\ref{tilted}. We adopted a $\Delta r$=0\farcs3, which corresponds to the de-projected resolution of our observations in the galaxy plane. The center position adopted in the fitting was fixed to correspond to the center of the nuclear spiral structure (see Table~\ref{prop}). In the right panel, we display the residuals after subtracting the tilted-model ring to the velocity field. As can be seen in Fig.~\ref{tilted}, the tilted-ring  represents  the observed velocity field very well, with small amplitude values in the residuals ($\pm$30\,km/s), indicating that mean velocity field is well described by rotation.

\begin{figure}
 \resizebox{\hsize}{!}{\includegraphics{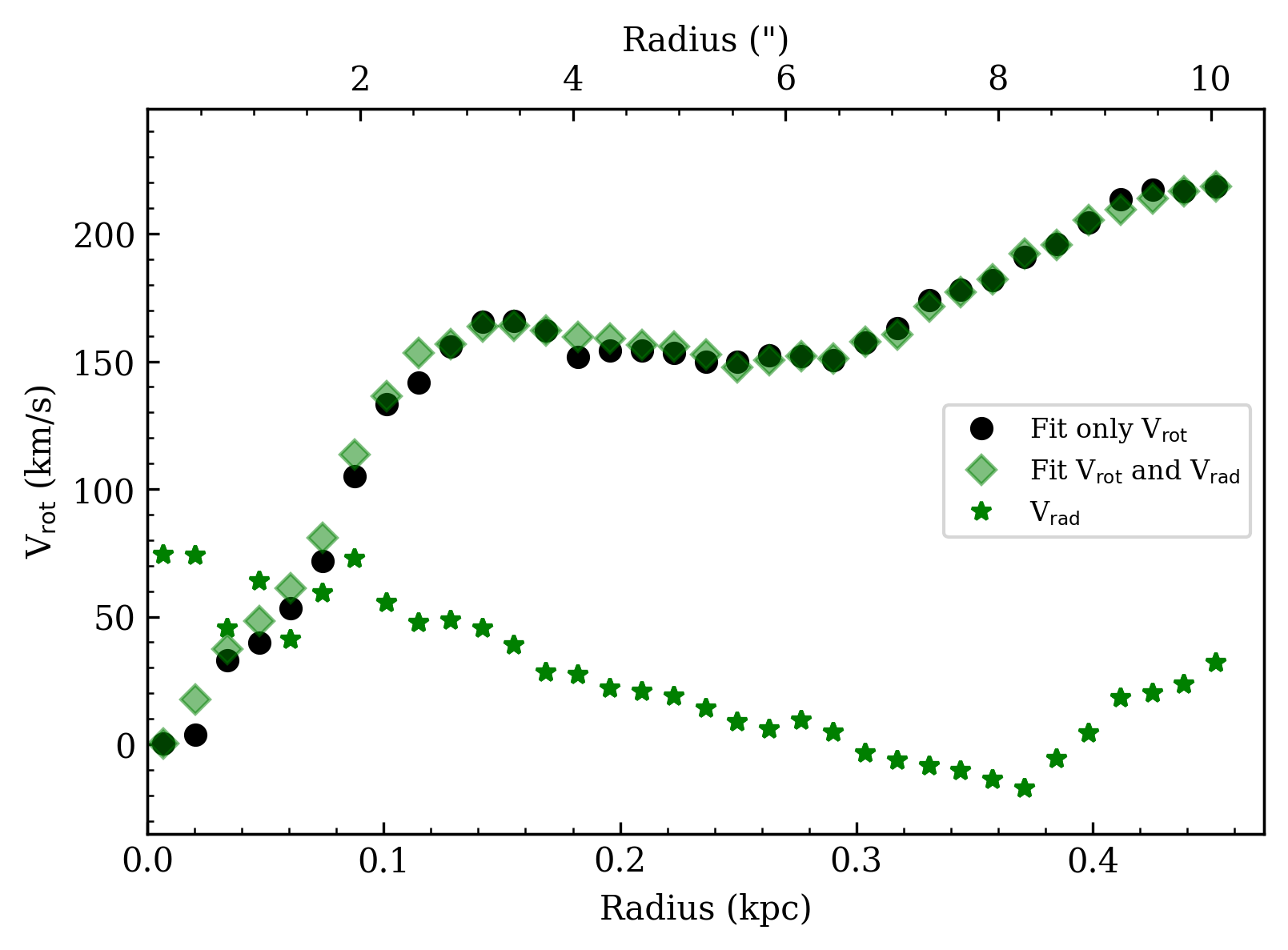}}
\caption{CO(3-2) rotation curve of NGC\,1808 obtained with the \textsuperscript{3D}BAROLO fitting. The black circles show the best fit for only circular motion and the rotation curve including radial velocities is indicated with green diamond symbols, together with corresponding $\rm v_{\rm rad}$ (green stars).}
\label{vel_rot}
\end{figure}

We also tested an additional method to derive the CO kinematics, using the 3D-Based Analysis of Rotating Objects from Line Observations ( \textsuperscript{3D}BAROLO) software from \citet{barolo}, which performs a 3D tilted-ring modeling of the emission line data-cubes to derive the parameters that better describe the kinematics of the data. We ran \textsuperscript{3D}BAROLO on the CO(3-2) data-cube in order to investigate non-circular motions since the code allows us to infer radial velocities, $\rm V_{\rm rad}$, in the fit of the rotation curves. The first time we ran the code, we allowed the PA, inclination, and fitting of the rotation curve and the velocity dispersion to vary, without including radial motion and only fixing the center as the nuclear spiral central position according to the coordinates listed in Table~\ref{prop}. We found a good agreement between the PA and inclination derived from \textsuperscript{3D}BAROLO and in the 2-D tilted ring approach described above; therefore, we fixed the PA=314$\rm ^\circ$ and i=57$\rm^\circ$ to the mean values between them.
The code was performed again now fixing the PA and $i$ to these values, and the resulting rotation curve is shown in Figure~\ref{vel_rot}. When we include radial velocities in \textsuperscript{3D}BAROLO, the best fit displayed in Figure~\ref{vel_rot} reveals radial components of order $V_{\rm rad}\sim$40-75\,\kms in the central 100\,pc. In this region, we detected the nuclear spiral and the radial velocities, which may be due to the non-circular motion that is expected for elliptical orbits in barred potentials. The discussion of these radial velocities in terms of an inflow are discussed in Section~\ref{inflow}.

\begin{figure}
 \resizebox{\hsize}{!}{\includegraphics{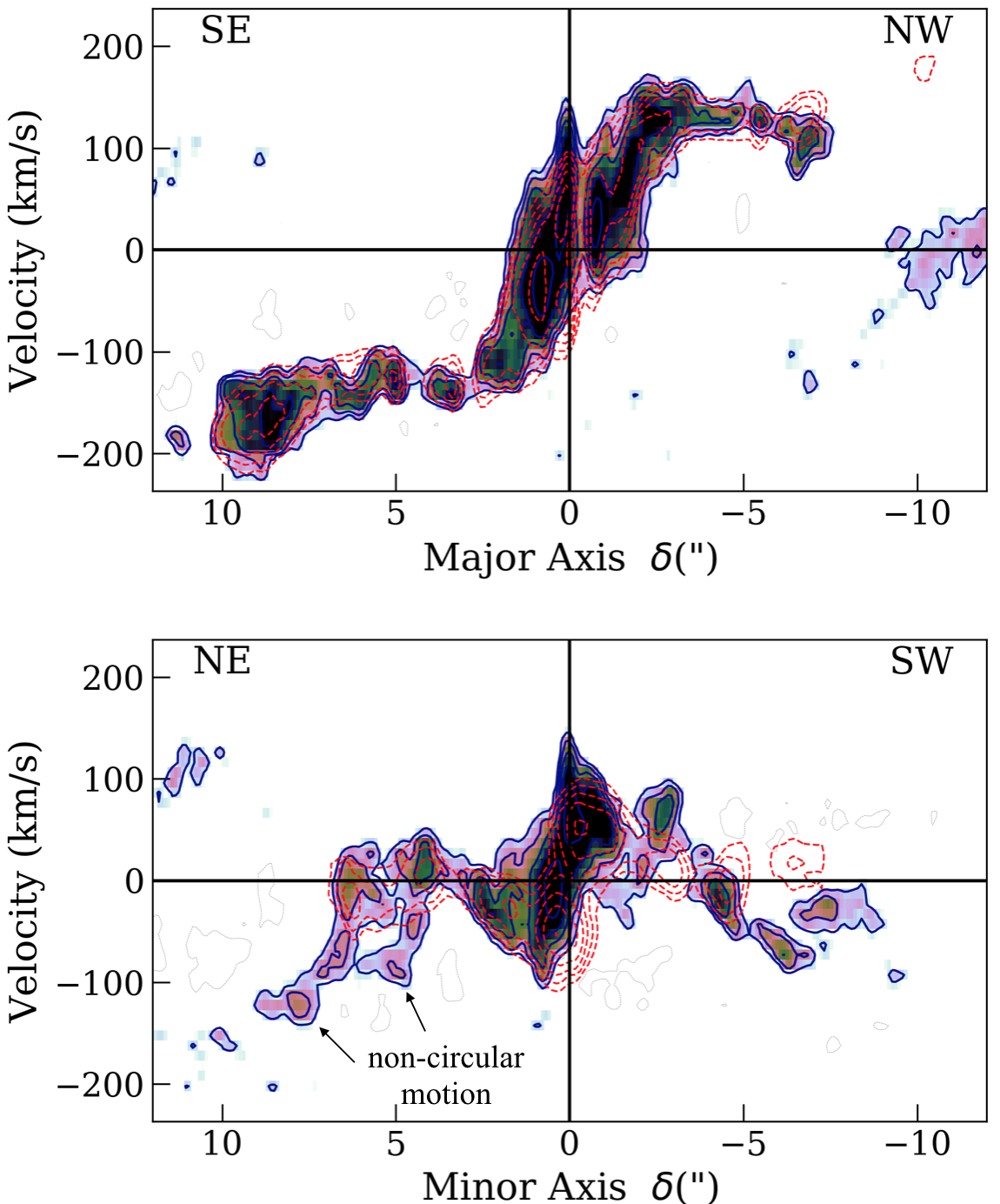}}
\caption{Position-velocity diagrams of NGC\,1808  along the major axis at a PA=314$^\circ$ (\textit{top panel}) and minor axis (PA=224$^\circ$, \textit{bottom}). We used a 0\farcs3 slit width. The red dashed lines are the best fit from  \textsuperscript{3D}BAROLO. The contour levels are shows as $2^n\times\sigma_{\rm rms}$, for n=1,2,...,7. Negative contours at 2$\sigma_{\rm rms}$ are shown in gray.}
\label{pv}
\end{figure}

We investigate the position-velocity diagrams (PVDs) along the major (PA=314$^\circ$) and minor axis (224$^\circ$) of NGC\,1808 (Fig.~\ref{pv}) to detect a possible outflow feature that would correspond to the polar dust lanes. Our CO(3-2) observations along the minor axis reveal that the bulk of the molecular gas corresponds to the nuclear spiral arms, a source of important non-circular motions. Within the central arcsecond, a high-velocity gas (100-200~\kms) is present, associated to the spiral arms and with the same morphology. Outside a radius of one arcsec, the velocities are mainly due to circular rotation and some perturbations from co-planar streaming motions along the spiral arms. In the very center (6\,pc), there is a velocity gradient that is likely due the fact that the kinematic is perturbed because of the decoupling of the torus, in which we observe a change in the PA. We discuss this misalignment in Section~\ref{misal}. 

\citet{salak16,salak17} presented evidence of a molecular gas outflow seen in CO(1-0) with ALMA from the nuclear starburst region (r$\lesssim$250\,pc) of NGC 1808 that could be only detected in the PVDs. The outflow has a maximum de-projected velocity of $v_{\rm out}\sim$180\,km/s and mass outflow rate between 1-10 M$_\odot$/yr, comparable to the total star formation rate in the starburst 500\,pc region, that corresponds to SFR$\sim$5M$_\odot$/yr in the ``hot spots.” In the PVD along the minor axis in our CO(3-2) observations, the region 5-10\arcsec NE also shows velocities of order of $\sim$-100\,km/s that cannot be fitted considering purely circular velocities. The non-circular motion can indeed be attributed to a molecular outflow, if we consider that NE side is the far side, supported by the presence of dust lanes in the SW side, as seen in the optical image in Figure~\ref{cgs}. This assumption is based on the trailing pattern of the outer winding spiral arms and, therefore, the negative velocities along the NE direction could not be explained by co-planar inflowing gas in the disk plane; in this case an outflow scenario is most possible. For a co-planar outflow, the de-projected outflow velocity is $v_{\rm out}=v_{\rm non-circ}/sin(i)\approx$120\,km/s and in the case the outflow is perpendicular to the disk, $v_{\rm out}=v_{\rm non-circ}/cos(i)\approx$180\,km/s. Considering the outflow has a certain geometry with an angle $\alpha$ between the outflow direction and the line of sight, the outflow velocity is $v_{\rm out}=v_{\rm non-circ}/cos(\alpha)$. Additional evidence supporting the outflow scenario comes from the increased velocity dispersion in the NE part of the disk, in which the dispersion is as high as $\sim$60\,km/s (Figure~\ref{moms}). Therefore, our results are compatible with the outflow previously proposed by \citet{salak16} and suggest that the molecular outflow is probably connected to the kpc-scale superwind seen in optical wavelengths.

\subsection{Dense gas: HCO$^+$, HCN, and CS emission}\label{dense}

\begin{figure}
 \resizebox{\hsize}{!}{\includegraphics{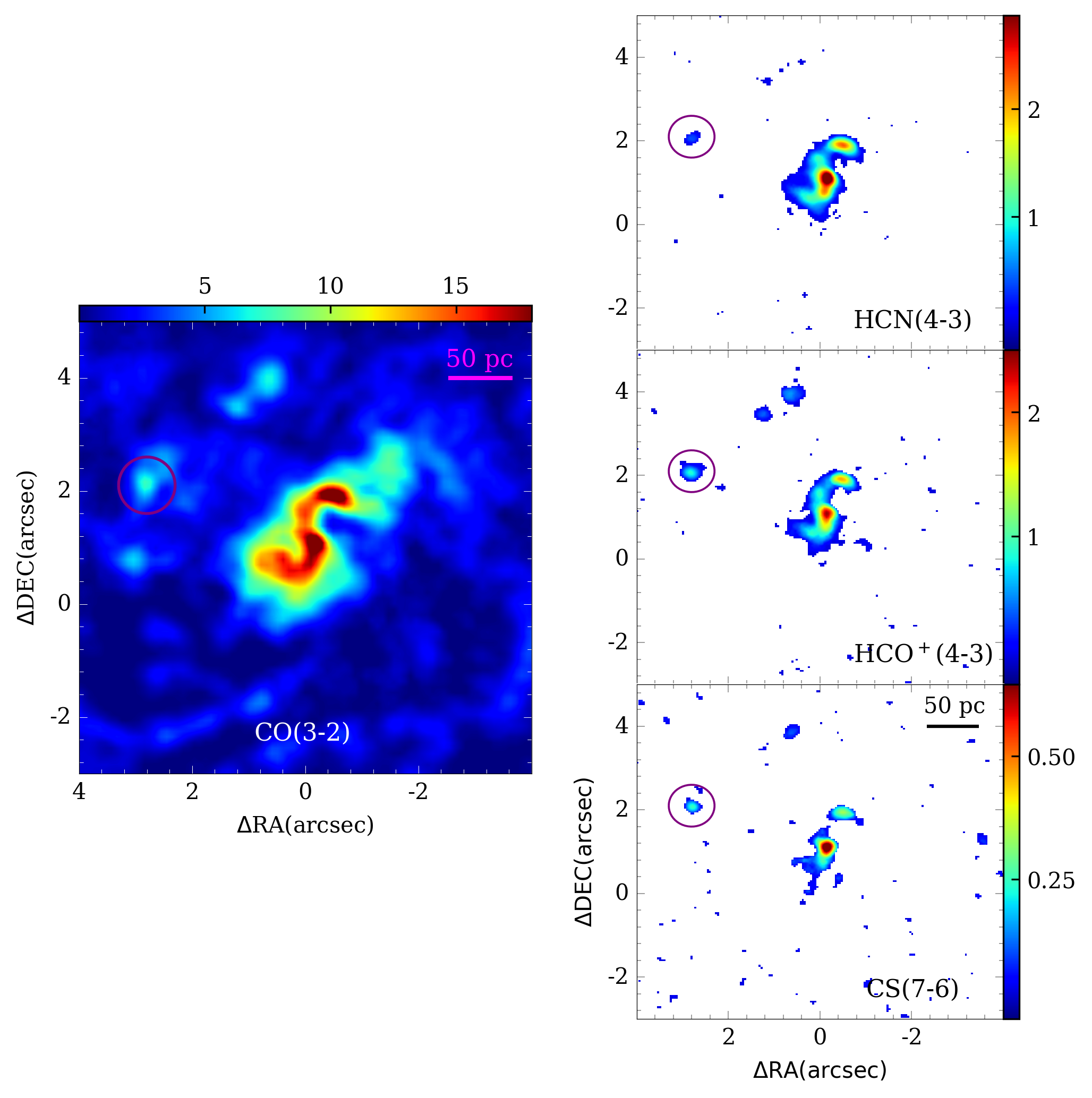}}
\caption{The nuclear spiral structure detected in CO(3-2), HCN(4-3), HCO$\rm ^+$ and CS(7-6) emission. \textit{Left:} Zoomed 8\arcsec$\times$8\arcsec\ region of CO(3-2) intensity map for NGC\,1808 of Figure~\ref{moms}. \textit{Right:} the integrated intensity maps of the dense gas tracers HCN(4-3), HCO$\rm ^+$(4-3) and CS(7-6) in the top, middle and bottom panels, respectively, in the same region. The purple circle indicates the position of a clump detected in the HCN, HCO$\rm ^+$ and CS.}
\label{densegas}
\end{figure}

As mentioned before, the tuning configuration of the observations allowed us to observe the high density gas tracers HCN(4-3), HCO$\rm ^+$(4-3), and CS(7-6). The detection of these transitions is shown in Figure~\ref{densegas}. The two-arm nuclear spiral traced by CO(3-2) is also visible in HCN(4-3), HCO$\rm ^+$(4-3), and CS(7-6) maps.

Regarding the ambiguous Starburst/AGN nature of NGC\,1808, the presence of a weak AGN is still debatable. Based on the $\rm H\alpha$/$[$N\textsc{ii}$]$=0.94$\rm \pm$0.05 line ratio classification of \citet{veron86}, NGC\,1808 is classified as a Seyfert galaxy. Ginga X-ray data (1.5–37\,keV) by \citet{awaki93} show evidence of an obscured AGN and it is attributed as Seyfert\,2. However, other observational evidence points to the starburst nature of the excitation mechanism in NGC\,1808 \citep{phi93,dop15}. \textit{XMM-Newton} and \textit{Chandra} observations by \citet{Jimenez-Bailon2005} indicate the co-existence of a starburst component along with a LLAGN contribution in the nucleus. 

\begin{figure}
 \resizebox{\hsize}{!}{\includegraphics{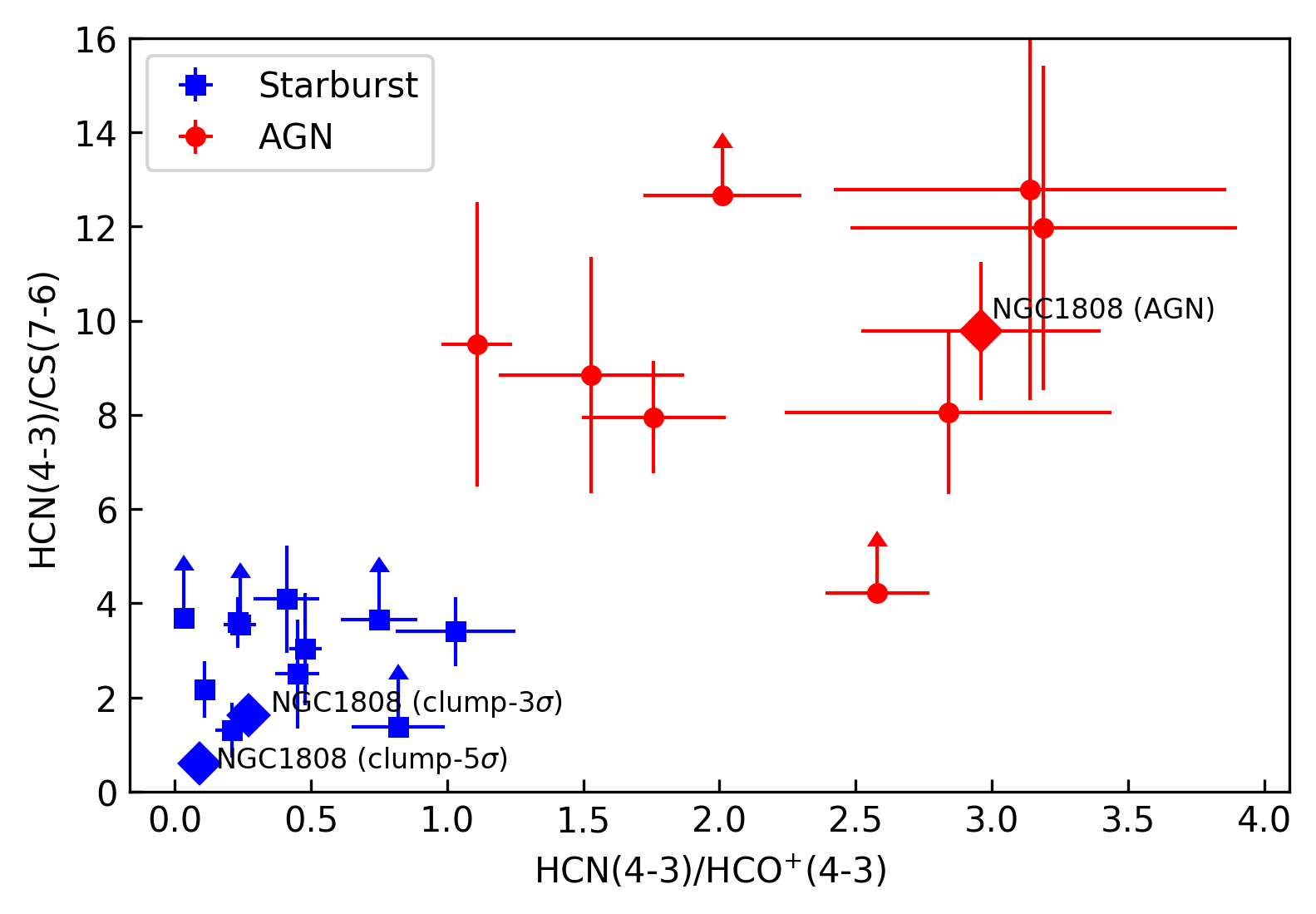}}
\caption{Submillimeter-HCN diagram proposed by \citet{izumi16} for the high-resolution observations (spatial resolution $<$500\,pc) using the line intensity ratios R$_{\rm HCN/HCO^{+}}$ and R$_{\rm HCN/CS}$. Red circles represent the AGNs and the blue squares indicate the SB galaxies. We include the line ratios of NGC\,1808 (diamonds) measured in the central spiral, called here as "AGN" and in a clump detected $\sim$150\,pc north-west of central position in all the dense tracers.}
\label{submm}
\end{figure}

Thanks to our high-resolution ALMA observations, we were able to disentangle the emission coming from the nuclear region within the spiral trailing feature observed in the central $\sim$50\,pc and the contribution from a SF clump observed in all the molecular tracers at $\sim$150\,pc from the nucleus, indicated in Figure~\ref{densegas}. We measured the line intensity ratios R$_{\rm HCN/HCO^{\rm +}}$ and R$_{\rm HCN/CS}$ in the two regions: in the nucleus (or AGN) and in the SF clump. For the latter, we computed two different intensity ratios, one using line emission above 3$\sigma$ per channel and the other considering in each channel only the emission higher than 5$\sigma$. The values for the line ratios are listed in Table~\ref{tab:submm} and presented in the submillimeter-HCN diagram (Fig.~\ref{submm}) proposed by \citet{izumi16}.

The diagram suggests an enhanced HCN(4-3)/HCO$^{\rm +}$(4-3) and HCN(4-3)/CS(7-6) integrated intensity ratios in circumnuclear molecular gas around AGN compared to those in starburst galaxies \citep[submillimeter HCN-enhancement,][]{izumi16}. The starburst and AGN points (blue and red symbols, respectively, in Figure~\ref{submm}) are taken from the ``high-resolution" sample of \citet{izumi16}, in which the observations have resolution in the range of 5 to 500\,pc and the difference between AGN and starburst is clear in the diagram. As highlighted by the authors, high spatial resolution observations are crucial for disentangling the contribution from the circumnuclear SF regions and AGN, making the submillimeter-HCN diagram a powerful diagnostic for identifying LLAGN in the case of composite systems. The ratios in the nucleus of NGC\,1808 are compatible with those found for AGN whereas the SF clump presents lower ratios, which is typical of starburst galaxies. Therefore, we do find that the nuclear region of NGC\,1808 presents excitation conditions that are typical for X-ray dominated regions (XDRs) in the vicinity of AGN, supporting the notion and some evidence of the presence of an AGN in the nucleus of NGC\,1808.

\begin{table}
\caption{Line ratios: R$_{\rm HCN/HCO^{+}}$ and R$_{\rm HCN/CS}$}          
\label{tab:submm}      
\centering                                      
\begin{tabular}{c c c}         
\hline\hline                       
Region & R$_{\rm HCN/HCO^+}$   & R$_{\rm HCN/CS}$  \\  
\hline                                  
 AGN & 2.96$\pm$0.44   &   9.78$\pm$1.47   \\
 Clump (3$\sigma$)& 0.27$\pm$0.04 & 1.63$\pm$0.24 \\
 Clump (5$\sigma$) & 0.09$\pm$0.01 &  0.59$\pm$0.09 \\
\hline        
\end{tabular}
\\
\tablefoot{Ratios of the HCN, HCO$\rm^+$ ans CS lines in the CND region and in the clump shown in Fig.~\ref{densegas}. \\
}
\end{table}

\section{Discussion}

\subsection{Ionized and warm molecular gas}

\begin{figure*}
\centering
\includegraphics[width=17cm]{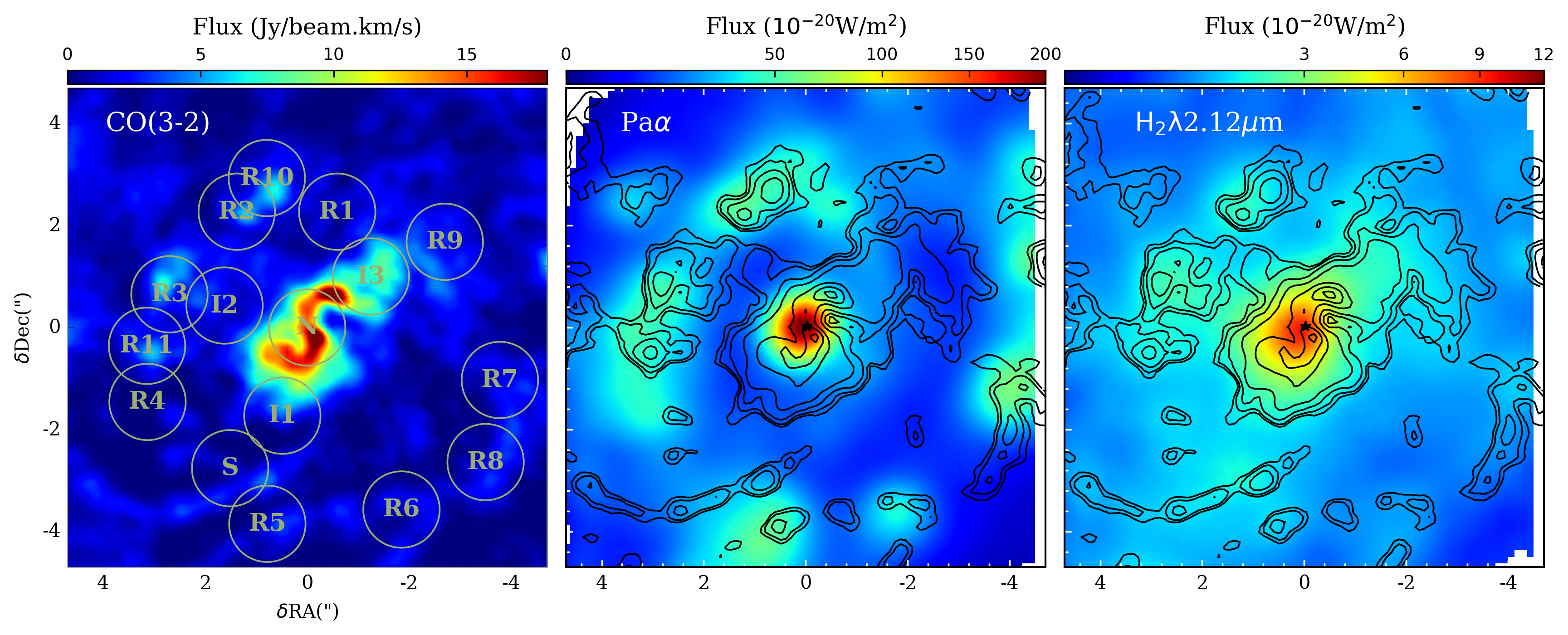}
\caption[SINFONI observations and CO(3-2).]{Comparison between the CO(3-2) emission in the central 9\arcsec$\times$9\arcsec\ (left panel), with the Pa$\alpha$ (\textit{middle}) and H$\rm_{2}\lambda 2.12\mu m$ (\textit{right}) emission in the 8\arcsec$\times$8\arcsec FoV of the SINFONI observations \citep{bus17}. The CO contours are overlaid in the NIR images. The black star represents the AGN position listed in Table~\ref{prop}. Left panel: 15 regions are identified, R1--R11, I1--I3 and N, to compute the Kennicutt-Schmidt relation of Fig \ref{KS-regions}.}
\label{sinf}
\end{figure*}

In order to compare the CO(3-2) morphology with the ionized material and the warm molecular gas, we superposed onto Figure~\ref{sinf} the CO contours onto the NIR maps obtained with SINFONI of Pa$\alpha$ and H$\rm_{2}\lambda 2.12\mu m$ presented in \citet{bus17}.  The SINFONI observations have a FoV of 8\arcsec$\times$8\arcsec and a seeing resolution of 0.4\arcsec, with a pixel of 0.125\arcsec, larger than this ALMA configuration (0.2\arcsec beam, 0.098\arcsec/pixel). There is a remarkable resemblance between the CO emission and the ionized and warm molecular gas along the SF ring at $\sim$4\arcsec radius. This SF ring is better traced by radio continuum emission \citep{col94}, Pa$\alpha$, and also the shock tracer [Fe\textsc{ii}] than by the excited H$\rm_{2}$ emission \citep{bus17}. It shows hot spots of star formation, corresponding to molecular cloud associations, of typical sizes 50\,pc, composed of several smaller molecular clouds. 

According to \cite{salak2018}, all molecular clouds along the SF ring have density and high HCN(1-0)/HCO$\rm^+$(1-0) line ratio (of order $\sim$1), possibly indicating that the excitation mechanisms are from shocks, free–free emission from H\textsc{ii} regions, and synchrotron emission from SNRs. This ratio even is higher towards the nucleus, with HCN/HCO$\rm^+ \sim$1.5, which is consistent with the scenario of a composite starburst plus AGN \citep{privon15,salak2018}. In addition, they also detected SiO(2-1), a shock tracer, only around the nuclear region of NGC~1808.

\subsection{Resolved Kennicutt-Schmidt relation}

\begin{figure}
\centering
\includegraphics[width=8.5cm]{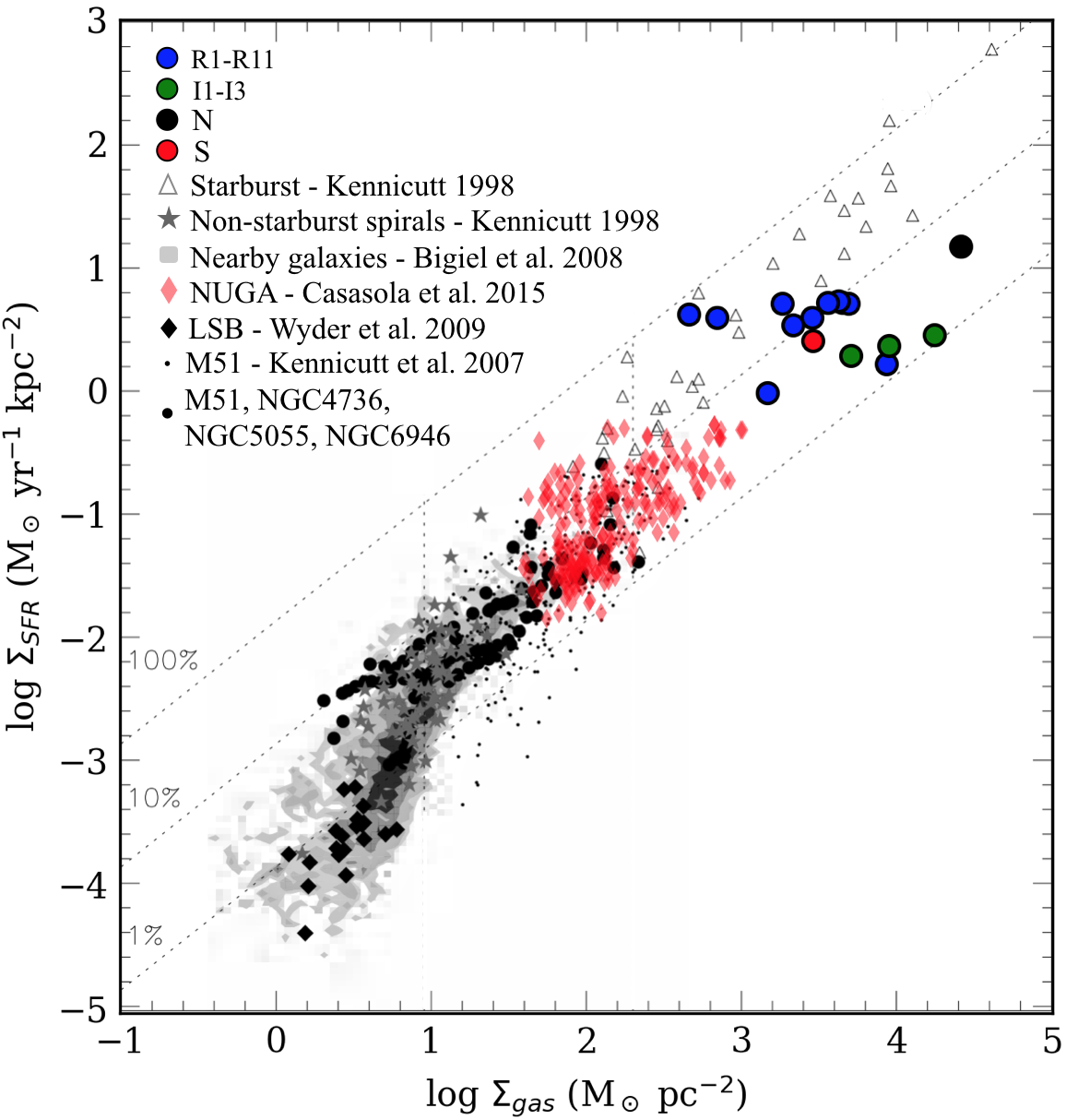}
\caption{Relation between the SFR determined from Pa$\alpha$ and gas surface densities, in the 16 regions delimited in Figure
\ref{sinf}. The nuclear region (N) is indicated by the big black circle, the region along the circumnuclear ring (R1-R11) in blue, the south region (S) in red, and in intermediate (I1-I3) regions in green. The NGC~1808 data points are compared to the KS relation from different surveys: the resolved KS law in nearby galaxies by \cite{Bigiel2008} in gray; the NUGA sample by \citet{vivi15}, shown as red diamonds; global measurements of Starburst galaxies (open triangles) and non-Starburst spirals (gray stars) by \cite{kenni98}; \cite{kenni07} measurements in individual apertures of M~51 (black dots); and radial profiles in the nearby spiral galaxies M~51 \citep{schuster07}, NGC~4736 and NGC~5055 \citep{wong02} and NGC~6946 \citep{cros07}, indicated as small black circles. Figure adapted from \cite{Bigiel2008}.}
\label{KS-regions}
\end{figure}

Since we have high spatial resolution in the galactic center, both in the star formation rate and gas surface density, it is possible to probe the Kennicutt-Schmidt (KS) resolved
relation, even below the scale of $\sim$ 500pc, where the H$\alpha$ and CO emission depart from each other \citep{schruba2010,kruijssen2018}.
The regions considered for the comparison are indicated in the left panel of Figure \ref{sinf}. The obtained relation is presented in Figure \ref{KS-regions}.

\begin{figure}
\centering
\includegraphics[width=8.5cm]{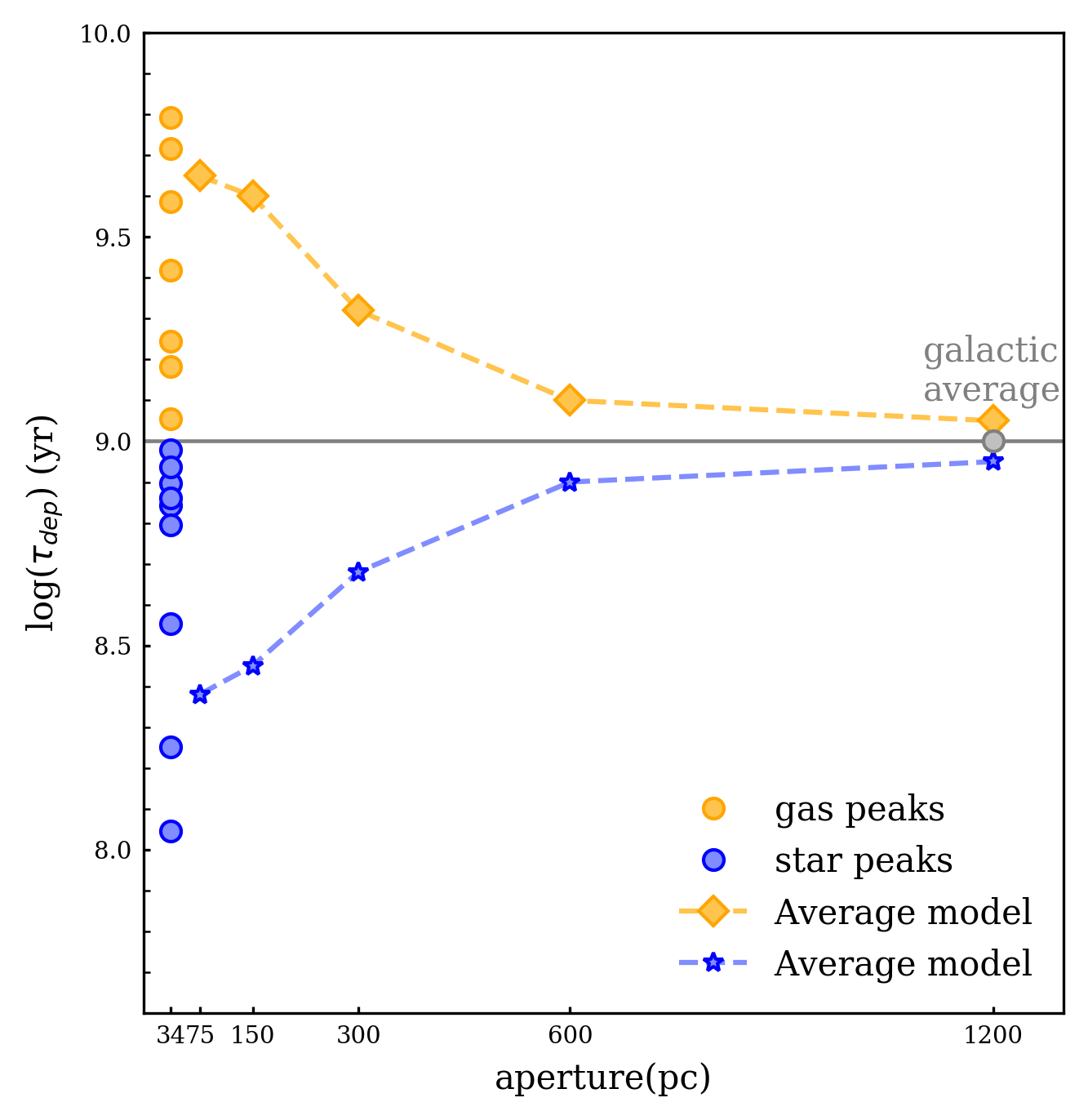}
\caption{Depletion time, that is, the ratio between gas and SFR surface densities,
as a function of the aperture, or region in which it is computed. This defines a minimum
region size of $\sim$ 500\,pc, to apply the resolved KS relation. The average star and gas peaks models are from \cite{schruba2010}.}
\label{unprin}
\end{figure}

             To better compare the ionized and molecular gas distribution, we compute the surface densities of 
             star formation rate (SFR) and molecular gas to build a resolved Kennicutt-Schmidt (KS) diagram. We first considered the same region of star formation identified by \cite{bus17}, although we know that they are biased towards H\textsc{ii} regions, and then consider some peaks of the CO(3-2) emission, that are not remarkable in SFR, namely, R8, R9, R10, R11, and I3 in Figure~\ref{sinf}. We applied the apertures used by \cite{bus17}, where the radius of 0.75\arcsec = 34\,pc, and we followed with the same sizes, for the sake of comparison. In the cited work, they derived the cold gas mass from the warm H$_2$ gas detected at 2.12$\mu$m, with a conversion factor between warm and cold molecular hydrogen given by the ratio of $M_{\rm H_2(cold)}/M_{\rm H_2(warm)}$=0.3-1.6$\times$10$^6$, which is highly uncertain. The SFR in all the regions shown in Figure~\ref{sinf} were computed using the integrated flux of the Pa$\alpha$ emission line in each region, corrected by the \citet{calzetti00} extinction law and applying the calibration of \citet{pan03}:
             \begin{equation}
                 SFR_{\rm Pa\alpha}=\frac{L(Pa\alpha)}{1.98\times10^{33}W} \rm M_{\rm \odot} yr^{-1}
             ,\end{equation}
             where $L(Pa\alpha)$ is the luminosity of the Pa$\alpha$ emission line in units of Watts. 
             
             To compute the molecular masses from our CO(3-2) cube, we integrated the fluxes in each aperture and applied a Milky Way-like conversion factor $\alpha_{\rm CO}=4.36$ and an intensity ratio of $r_{\rm31}=$0.27 \citep{bol13}. 
             The molecular masses were corrected from the 45\% missing flux, as discussed in Section~\ref{miss}, and all SFR and $M_{\rm H2}$ surface densities were corrected to the galaxy inclination ($i=$57$^\circ$). The resulting KS diagram is shown in Figure \ref{KS-regions}. We compare our findings with the work from \citet{Bigiel2008} that studied the star formation laws at sub-kpc scales in a sample of nearby galaxies and the work of \cite{vivi15} using HST and CO observations of the NUGA sample. All our points lie in the KS relation, although some of them present higher values of gas and SFR surface densities than the ones presented in \cite{Bigiel2008} and \cite{vivi15}. Most of the regions along the ring (blue points in Fig.~\ref{KS-regions}) are located at the upper part of the KS relation, indicating high SF efficiencies, similar to Starburst galaxies \citep[open triangles in Figure~\ref{KS-regions},][]{kenni98}. On the other hand, the intermediate regions I1-I3 (marked as green circles) and R9 in the ring all show lower star formation efficiencies.
             
             A possible explanation could be due to the bias of the apertures selection according to the stars or gas peaks. In order to address the effects of the resolution and aperture selection, in Figure~\ref{unprin}, we show two series of points: the bias in finding more SFR when the selection is done on the optical (star peaks) or more gas surface density when the selection is done on the molecular clouds (gas peaks) is quite clear. The regions I1-I3 and R9 are defined as the gas peaks and correspond to the region with lower SF efficiencies in the KS relation. The result of the bias selection trend is translated into the depletion time, $\tau_{\rm dep}=\frac{\Sigma{gas}}{\Sigma_{\rm SFR}}$, and we found that the gas peaks show higher depletion times than the star peaks. This effect has been well described already by \cite{schruba2010}, and in more details= by \cite{kruijssen2018}: at a small scale, typically lower than  500\,pc, we expect that the young stars and their surrounding ionized gas dissociate from the molecular association where they were born. From this relation, it could be possible to derive several quantities, such as the star formation  efficiency at the giant molecular cloud-scale,  the  feedback  efficiency in terms of outflow  velocity and mass ejected, and its coupling efficiency to the molecular disk. However, we do not have enough regions to apply this method, in the small FoV of SINFONI (8\arcsec = 360\,pc) relative to the seeing resolution, thus, we postpone this study to the future when more extended observations become available. A more detailed analysis on the resolved KS law and star formation efficiencies for all galaxies in the sample presented in \citet{combes19} will be addressed in a forthcoming paper.

\subsection{Kinematic misalignment at the center of the nuclear spiral}\label{misal}

\begin{figure}
\centering
\resizebox{\hsize}{!}{\includegraphics[width=8.5cm]{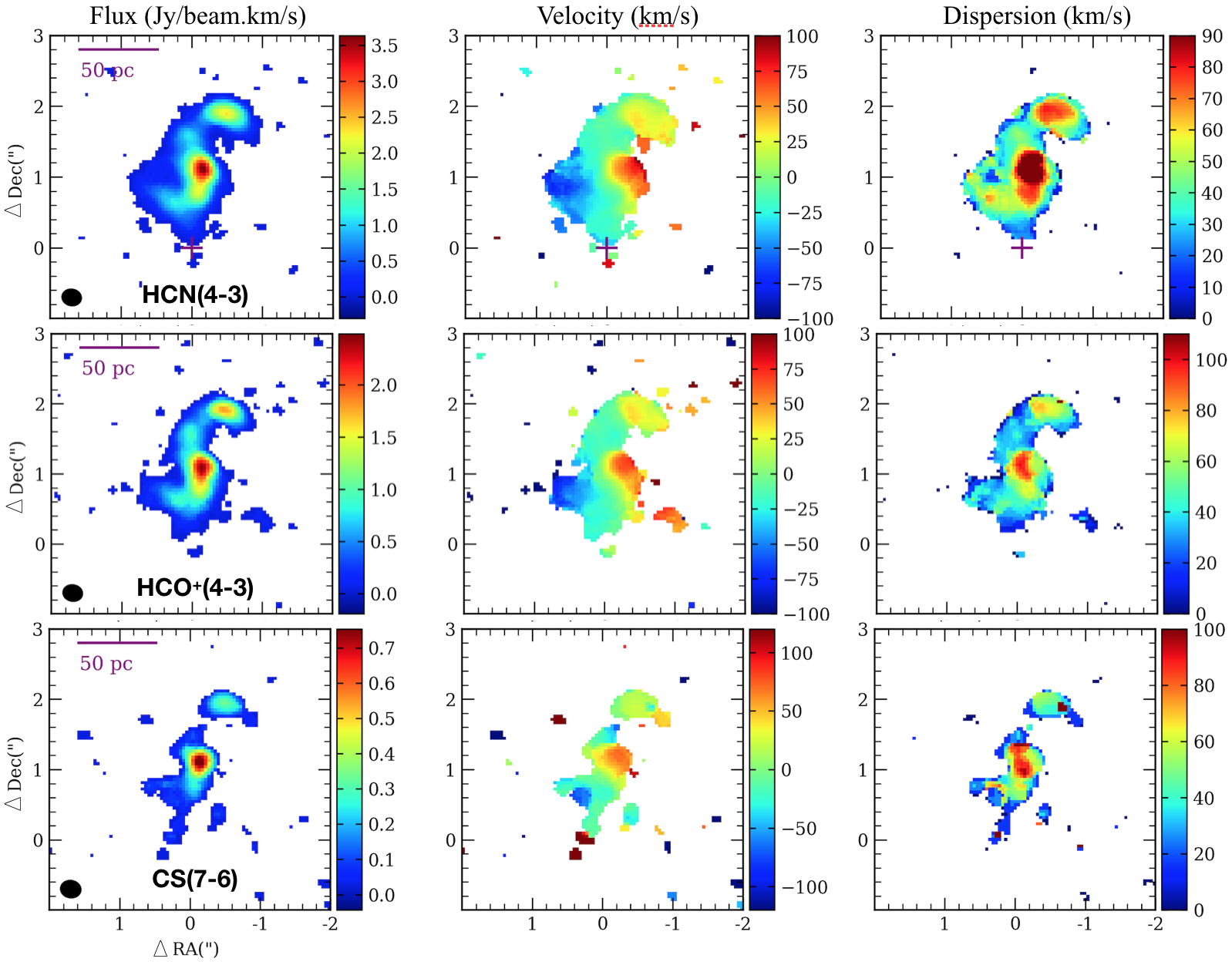}}
\caption[HCN(4-3), HCO$\rm ^+$(4-3) and CS(7-6) moment maps.]{The moment maps of the central emission of HCN(4-3), HCO$^{\rm +}$(4-3) and CS(7-6) (from \textit{top} to \textit{bottom}). The synthesized beam sizes are show in the bottom left corner of the zero-moment map of each line. The purple cross is the phase of the observations.}
\label{dense_moms}
\end{figure}

As shown in the previous section, with the CO(3-2) velocity field, we can see\ a tilt in the kinematic major axis
towards the nucleus. ALMA observations of neutral atomic carbon, [CI], presented in \citet{salak19} at a spatial resolution of 0.82\arcsec\ $\times$0.6\arcsec,\ show a remarkable resemblance to the CO(3-2) morphology and kinematics. They also reveal this tilted kinematics in the center. This decoupling kinematic feature is the indication of a nuclear disk and can be seen even in the dense gas tracers. In Figure~\ref{dense_moms}, the intensity and velocity moment maps of  HCN(4-3), HCO$\rm ^+$(4-3) and CS(7-6) are shown. Focusing on the center, the tilt becomes more apparent. The spiral arms tend to follow the same kinematics than the main disk. But towards the nucleus, there is a decoupled circumnuclear disk (or molecular torus). 
This nuclear disk does not have the same kinematical PA as the galaxy disk (314$^\circ$), but it is tilted by  about 30$^\circ$ towards 280$^\circ$. Such a kinematical decoupling is frequent and not unexpected \citep{santi19,combes19}.

\subsection{Inflowing of cold and warm molecular gas}\label{inflow}

The hot and cold gas, as well as the stellar velocity field, all show a very similar degree of rotation and, therefore, a simple residual map, obtained by subtracting the stellar velocity field from the velocity fields of both ionized gas traced by Br\,$\gamma$ and warm molecular gas (traced by H$_2$) emission, can provide us with an estimate of the non-rotational motions in the gas velocity fields  \citep{bus17}. The high spatial resolution (<50\,pc) of SINFONI observations show evidence of a nuclear two-arm spiral structure in the central 100\,pc (marked with solid lines in the residual maps presented in Figure~14 of \citet{bus17}). 

Assuming that they are located within the disk plane and that the near side of the disc is in the south-west,
 justified by the velocity field, and the trailing nature of the outer spiral arms, \citet{bus17} concluded that the residual spiral arms could correspond to streaming motions towards the centre. The authors also compared the spiral arms with the optical HST/F614W map, and the possible inflow motions coincide with dust features. 

The two-arm spiral structure detected in the residual maps of SINFONI observations reveal a remarkable coincidence with the nuclear spiral seen in our dense cold gas observations, giving additional support to the presence of the inflow of gas towards the nucleus of NGC\,1808. The quantification of the amount of gas possibly fueling the nucleus is discussed in the following sections.

\subsection{Lindblad resonances}

\begin{figure}
\centering
\includegraphics[width=8.5cm]{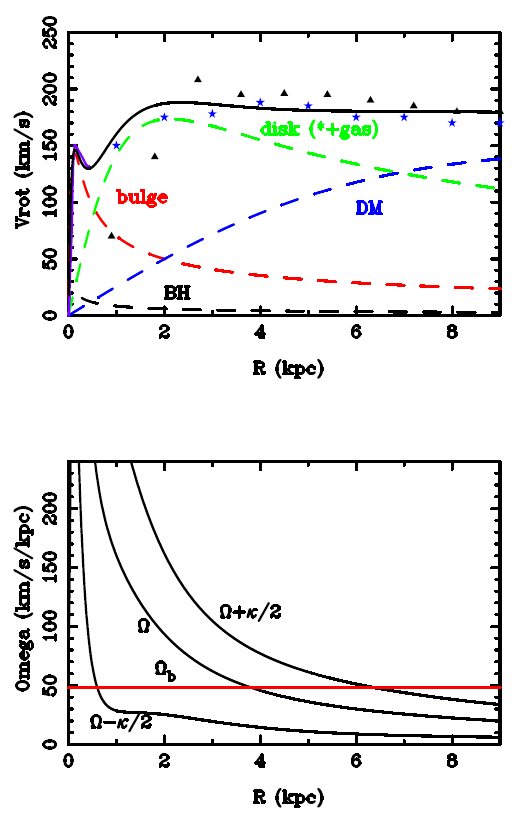}
\caption{Rotation curve model for NGC~1808. {\it Top}: An analytical model is built with five mass components: the stellar bulge and disk, the dark matter halo, the black hole, and the gaseous disk. The latter is included in the stellar disk. The contribution of each component is indicated as a dashed line. The blue symbols (stars) are adapted from the observations of H$\alpha$ by \cite{saikia90} (average of the NW and SE sides) and the black triangles from the HI observations
\citep{kor93}. Our CO rotation curve, corrected for bar perturbation, is indicated by the purple line. {\it Bottom}: The derived frequencies $\Omega$, $\Omega\pm\kappa/2$, indicating the Lindblad resonances, with the pattern speed of the bar highlighted in red.}
\label{Omegabar}
\end{figure}

NGC~1808 is a strongly barred late-type spiral galaxy, with typical dust lanes in the leading
edges of the bar, the latter being also delineated by a series of HII regions all along 6\,kpc 
\citep{kor96}. The 10\arcsec radius CO ring \citep{salak16,salak17} corresponds to a conspicuous 
ring in H$\alpha$, meaning that the gas is piling up in this ring and forms stars actively. This ring is likely the location of the inner Lindblad resonance. 
To check this, we made a model of the rotation curve to fit the observations of HI and H$\alpha$, adopting an inclination for the galaxy of 57$^\circ$. We combined the different
observations by \cite{saikia90}, \cite{kor93} and our work, adopting a common distance of D=9.3 Mpc (cf Fig. \ref{Omegabar}).
For the HI rotation curve, points of maximum velocity were measured on the PV diagram of \cite{kor93} every 20\arcsec, then the values from the two sides were averaged and divided by sin(inclination).

To fit the rotation curve, we consider several mass components. First, the baryonic ones are constrained by
the observations: the stellar bulge, better imaged in red colors, the stellar disk, and the gas disk,
with total masses
listed in Table~\ref{prop}. The black hole mass is estimated from the M-$\sigma$ relation, as discussed in 
\cite{bus17} and \cite{combes19}. Although the mass of the black hole is quite small, it is a point source and its sphere of influence is 14-36pc, which is important in this context \citep{combes19}. The gas mass is also a very small fraction of the baryonic mass, but it is extended and included in the stellar disk. As for the dark matter, it is required for the rotation curve to flatten at large radii, however, it is negligible in the center. The region we are interested in simulating has a radius of 14\arcsec=630\,pc. The dark matter component is the free parameter to better fit the rotation curve V(R) and derive
 all frequencies $\Omega$ and $\kappa$, which allow us to determine the location of Lindblad resonances. It is therefore important to have analytic density-potential pairs for these derivations. We use Plummer spheres for spherical components, bulge and dark matter, and Toomre disks for the 2D ones.
The parameters adopted for all mass components, reproducing the rotation curve, are displayed in Table \ref{tab:mass}. The dark matter mass is only the mass within 10\,kpc. Given the freedom and the degeneracy to distribute all mass components, within the uncertainties of the rotation curve, the parameters are not unique and just a realistic representation of the potential.

We have placed the corotation at $\sim$1.2 bar radius, as is usually found in numerical simulations and observations \citep{oneill2003}. There exists an inner Lindblad resonance
around the location of the CO/H$\alpha$ ring, at radius 10\arcsec = 450\,pc, which corresponds to the main ring
of molecular gas condensation.  

Inside this ILR, the SINFONI K$_s$ band measurements of \cite{bus17} have revealed a stellar nuclear bar, of 4\arcsec radius. Its orientation is comparable, albeit slightly inclined with respect to
the main bar of PA=145$^\circ$, itself with an angle towards the kinematic major axis of the galaxy of
PA=128$^\circ$, determined from stellar kinematics \citep{bus17}. Either the nuclear bar
is a prolongation of the main bar with the same pattern speed, just slightly tilted \citep{combes1994}, or a secondary bar has been decoupled from the main one, with a faster speed \citep{Friedli1993}.  We favor the first possibility, since the orbital precessing rate $\Omega-\kappa/2$ shown in Fig. 
\ref{Omegabar} is rather flat in the central kpc. The conditions to trigger the decoupling of a secondary bar, as shown by the simulations, are not yet fulfilled. The nuclear bar in the present configuration is the prolongation of the primary one; it can exert gravity torques on the gas, removing its angular momentum and driving the gas to the center. In the next section, we present the calculations of these torques from the observations themselves.

\begin{table}
\caption{Mass components}              % title of Table
\label{tab:mass}      
\centering                                      
\begin{tabular}{c c c c}         
\hline\hline                       
Component & Profile & R$_e$ & Mass  \\  
          & & (kpc)  &  (10$^9$ M$_\odot$) \\
\hline                                  
Bulge        & Plummer   & 0.1   &  1.31  \\
Disk         & Toomre    & 2.6   &  24.9  \\
Dark Matter  & Plummer   & 11.7  &  45.3  \\
Black Hole   & Keplerian &  --  &  0.016  \\
\hline        
\end{tabular}
\\
\tablefoot{
 The corresponding rotation curve is shown in Fig.~\ref{Omegabar}.
}
\end{table}

\subsection{Torque computation}\label{torques}
\begin{figure}
\resizebox{\hsize}{!}{\includegraphics{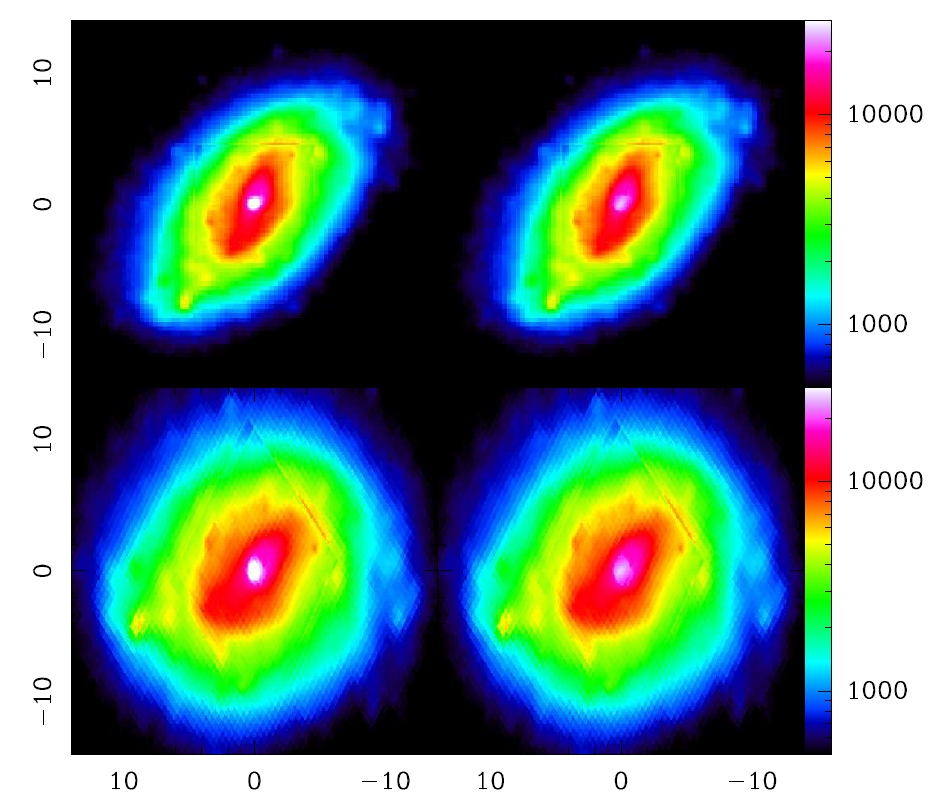}}
\caption{Combination of SINFONI and VISTA K-band images used in the computation of the stellar potential. {\it Top:} Map of the adopted K-band image, combining the SINFONI map in the central 8 \arcsec and the VISTA-K image at a large scale. From the total image at left, we subtracted a small spherical bulge in the right image. {\it Bottom:} Corresponding de-projected images, where the major axis is now horizontal. The axes are shown in \arcsec\  and the color bar is the same for the four images.}
\label{fig:nir-bsub}
\end{figure}

   To compute the stellar potential in the central kpc -- and, precisely, within the radius of 45\,pc --
   the radius of the gaseous nuclear spiral, one of the best images would have been
   the HST image. However, the red
   HST image is highly obscured by dust, up to the point that
   it is not possible to see the nuclear bar.
   Instead, we chose to use the SINFONI K$_s$ band image from \cite{bus17}, with a FoV
   of 8\arcsec$\times$8\arcsec\. To enlarge the FoV, we completed this image with
   the K image from the Visible and Infrared Survey Telescope for Astronomy (VISTA) of the Very Large Telescope (VLT), calibrating the borders to overlap with the same intensity.  The combined K-image used to obtain the potential is displayed in Fig. \ref{fig:nir-bsub} (top left).
   According to the method described in \cite{ane613}, we de-project the NIR image, after subtracting a spherical Plummer bulge component of a characteristic radius of 1\arcsec, to a face-on view (with PA= 311$^\circ$ and i=57$^\circ$). The K-image with bulge subtracted, together with its de-projected version, are also displayed in Fig. \ref{fig:nir-bsub}. The bulge subtracted here is somewhat smaller than the component required for the rotation curve; this is possible if the bulge is in fact partly flattened. We resampled both the derived potential and forces to the same grid as the CO(3-2) gaseous de-projected image, namely, with a pixel of 0.085\arcsec = 3.8\,pc. For this computation, we used the morphological center of the molecular nuclear spiral as our reference.

   Following the methodology presented in \cite{burillo2005} and \cite{combes1566}, we estimated the gravitational torques exerted by the stellar potential on the molecular gas. The gravitational potential is derived using a fast Fourier transform of the stellar mass distribution from the de-projected NIR image. In order to take into account the non-negligible thickness of the galaxy disk, we assume a thin disk of a scale ratio of $h_z/h_r$ =1/12, which is the typical scale ratio found for spiral edge-on galaxies \citep[e.g.,][]{barna92,biz09}. We neglected the contribution of dark matter within a 1\,kpc radius (the extent of this computation).  The final calibration was obtain using a constant mass-to-light ratio of M/L=0.95, which is necessary to retrieve the observed CO rotation curve.

\begin{figure}
\resizebox{\hsize}{!}{\includegraphics{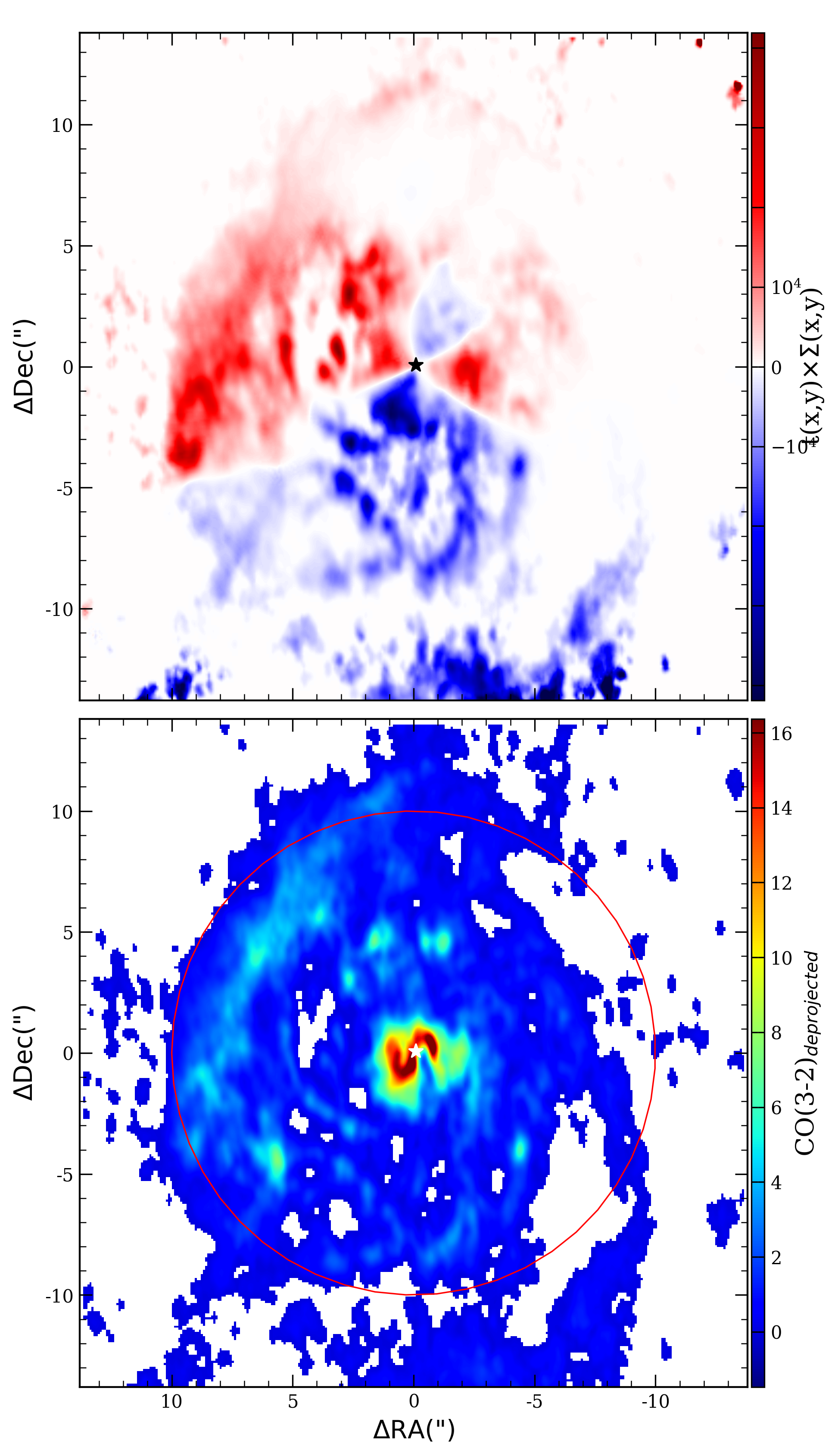}}
\caption{De-projected maps of the gravitational torque and the CO(3-2) emission. {\it Top:} Map of the gravitational torque,  t(x,y)~$\times$~$\Sigma$(x,y), in the center of NGC\,1808. The torques change sign as expected in a four-quadrant pattern (or butterfly diagram). The orientation of the quadrants follows the nuclear bar's orientation. In this de-projected picture, the major axis of the galaxy is oriented parallel to the horizontal axis. {\it Bottom:} De-projected image of the CO(3-2) emission, at the same scale and with the same orientation, for comparison. The ILR is indicated as a red circle.}
\label{fig:torq1}
\end{figure}

   The tangential forces per unit of mass ($F_x$ and $F_y$) were calculated as the derivatives of the potential and the torques at each pixel, $t(x,y)$, were then computed by $t(x,y) = x~F_y -y~F_x$. The product of the torque and the gas density $\Sigma$ at each pixel allows us to derive the net effect on the gas and $t(x,y)\times\Sigma(x,y), $ which is plotted in Figure~\ref{fig:torq1}, together with the de-projected CO map. In the top panel, the map of the gravitational torques shows a butterfly diagram (four-quadrant pattern) in relation to the nuclear bar orientation (PA$_{\rm bar}$=128$^\circ$) with torques changing sign as expected due to the barred potential.
   
\begin{figure}
\resizebox{\hsize}{!}{\includegraphics{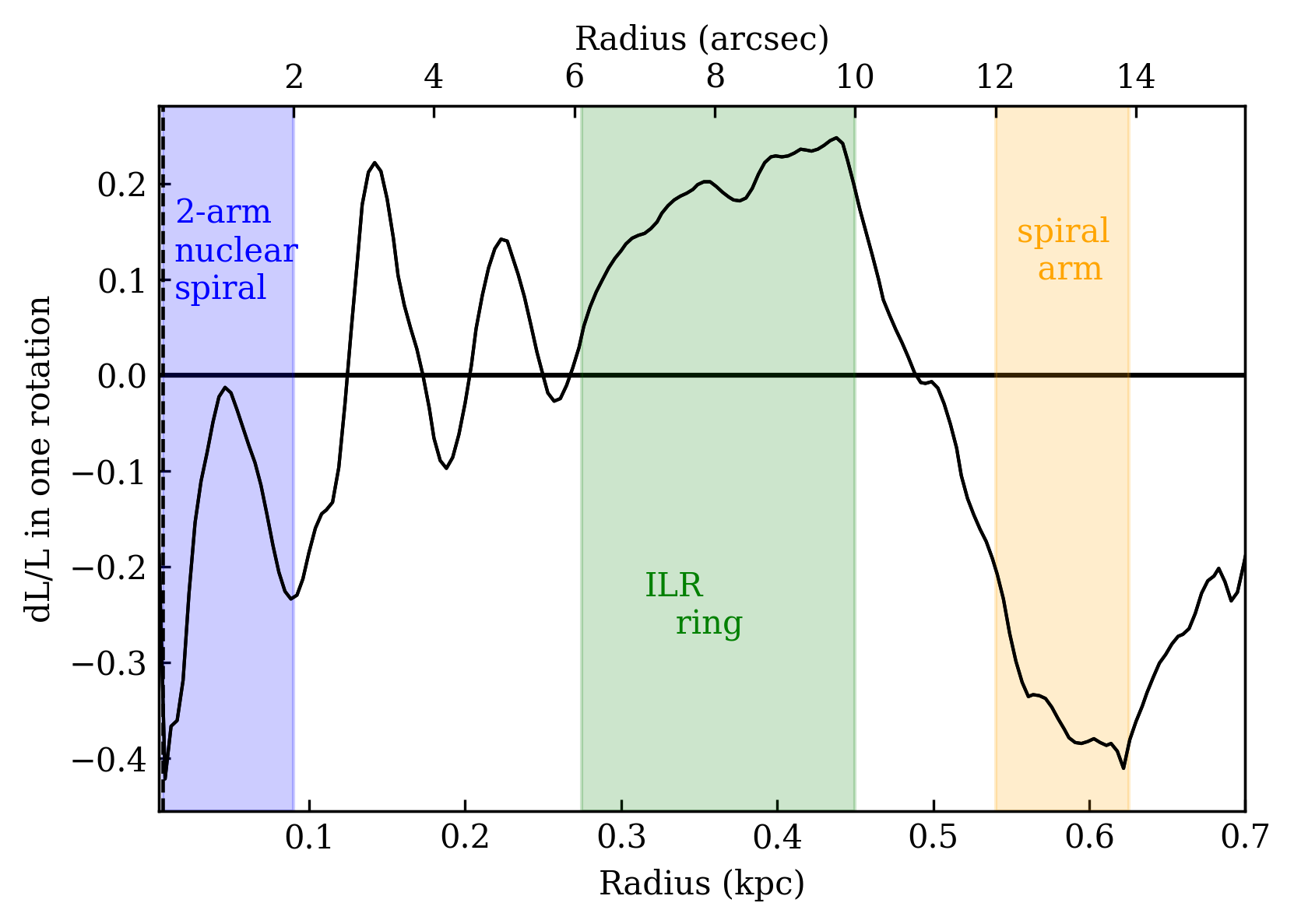}}
\caption{Radial distribution of the torque, quantified by the fraction of the angular momentum transferred from the gas in one rotation--$dL/L$, estimated from the CO(3-2) de-projected map. The vertical dashed line at 6\,pc radius delimitates the extent of the molecular torus, in this case, unresolved to match the resolution of the NIR image. The torque is negative inside the $\lesssim$100\,pc nuclear spiral and in the spiral arm at 
$\sim$600\,pc. Torques are positive at the ILR ring and in the filaments connecting the nuclear spiral and the ILR ring.}
\label{fig:gastor}
\end{figure}

   Through azimuthal averaging, we derived the effective torque at each radius, per unit mass. Dividing by the average angular momentum at this radius (from the rotation curve), we obtained the relative rate of transfer of angular momentum, as can be seen in Figure~\ref{fig:gastor}. The torque is negative in the nuclear spiral region, although it is relatively low. The gas is losing only 20-40\% of its angular momentum in one rotation. The torque is mostly positive inside the ILR, and negative outside the ILR region, meaning that the gas is somewhat maintained in the resonant ring, which explains the hot spots of star formation there. Only in the very central region, under the gravitational influence of the bulge and the black hole, the torque drives the gas towards the center.
    We note that neither the sign of the torques, nor the relative amplitude of angular momentum exchange per rotation, is dependent on the rotation curve adopted. Indeed, the amplitude of the torques depend on the asymmetry of the image and, therefore, the derived potential. In an axisymmetric disk, the gravity forces are only radial (F$_r$) and there is no torque. The relative amplitude of the torque depends on the ratio between the tangential to radial force, or $\alpha$ = F$_t$/F$_r$, and this depends only on the shape of the mass distribution. The result given in relative terms, dL/dt/L at a given radius, can vary as $\alpha$ V/R, where V is the rotation velocity at radius R, and, therefore, varies  as $\alpha$ /T, with T being the rotation period at this radius. The number of rotation for the gas to infall is independent of the adopted rotation curve, only when translated in physical time, the rate of exchange will follow the rotation timescale. 
   
   As explained in \cite{combes1566}, without any gravitational influence of a central mass, compact bulge and central black hole, we could expect the existence of two ILRs and a leading nuclear spiral structure inside the inner ILR, with a positive torque. However, most of the gas of the nuclear spiral is within the sphere of influence of the compact bulge and black hole, which is lower than 45\,pc according to the definition
   \citep{combes19}. The whole nuclear spiral is confined to a radius of 50\,pc and most of the central gas is imposing the trailing sense to the gas. This explains the negative torque that is driving the gas inwards.
    Figure~\ref{fig:gastor} allows us to quantify the intensity of the torque and the timescale of
    the gas infall. At a 100\,pc radius, the efficiency of the angular momentum transfer is $\sim$20\% and,  thus, we conclude that the gas is feeding the nucleus on a timescale of five rotations, meaning about $\sim$30\,Myr at the corresponding radii (at $r=100$\,pc, $v_{\rm rot}\sim 120$\kms and $t_{\rm orb}=6$\,Myr). In addition, we numerically integrated  the obtained curve $dL/L$ inside the 100\,pc radius, assuming that at each radius the orbit is quasi circular and the time for a gas cloud to reach the center is 100\,Myr -- if the potential is not varying over this timescale. On average, the inflowing timescale is in the order of $\sim$60\,Myrs.
    This short timescale, along with the evidence demonstrating that the active nucleus is directly fed by the bar due to a trailing
    nuclear spiral inside the ILR and within the sphere of influence of the black hole, has already been
    found in two other cases, namely, NGC 1566 \citep{combes1566}, NGC 613 \citep{ane613}. Other smoking-gun evidence of mechanisms fueling the galaxy center, albeit at a larger scale, has been obtained through other mechanisms, based on a decoupled secondary bar, as in NGC 2782 \citep{hunt2008}. This case corresponds to a more evolved stage, where the nuclear bar evolution has weakened the primary bar into an oval.

\section{Conclusion}
The high spatial resolution and high sensitivity of ALMA band 7 observations of the Seyfert2/starburst galaxy NGC~1808 allow us to analyse the distribution and kinematics of the molecular gas within the central few hundred of parcsecs, showing evidence of the feeding mechanism of the central SMBH.
In addition to a central point source, corresponding to the AGN, the continuum emission is extended, with a steep spectral index corresponding to synchrotron emission. We interpret this extended emission as coming from the star formation in the nuclear spiral and clumps in the pseudo-ring, of a radius 10\arcsec = 450\,pc.  

The CO(3-2) emission line cube revealed the morphology and dynamics of the gas inside the central 0.5\,kpc: inside the pseudo-ring, tracing the inner Lindblad resonance region, we identified a transition region with filaments that is linked to a nuclear disk of $\sim$ 200\,pc in radius. Inside this nuclear region, all the molecular lines reveal a very contrasted nuclear spiral of radius 1\arcsec = 45pc. This nuclear spiral is trailing, which is the configuration already encountered in two similar barred Seyferts, NGC 1566 and NGC 613 \citep{combes1566,ane613}. Trailing spirals are expected inside ILRs when the gas enters the sphere of influence of the central black hole; here this is combined with a compact bulge.

The dense gas tracers, namely, the HCN(4-3), HCO$\rm^+$(4-3), and CS(7-6) lines, are also clearly tracing  the nuclear spiral. We computed the gravitational potential due to a central nuclear bar that is clearly detected in the K-band image \citep{bus17}, with a similar orientation as the primary bar. The torques exerted on the molecular nuclear spiral by the barred stellar potential reveal that the gas is feeding the nucleus, on a timescale of five rotations -- or on an average timescale of 60\,Myr at these radii.

Inside the nuclear spiral, the molecular gas appears kinematically decoupled in a tilted circumnuclear disk of radius 0.13'' = 6\,pc, which is traced in CO(3-2) and also all dense gas tracers. This molecular disk or torus has HCN and HCO+ line ratios compatible with an excitation from the AGN. According to the distance from the center, the excitation changes progressively from AGN-like to starburst-like.

We searched for a possible molecular outflow in the nucleus since this galaxy is known to host a large-scale superwind \citep{salak16,salak17}. The kinematics in the center is indeed perturbed within a radius of one arcsec (45\,pc), showing a velocity gradient on the minor axis. These velocity perturbations follow the nuclear spiral morphology and are interpreted as the associated non-circular motions. We cannot rule out a weak AGN-induced outflow; however, the more energetic large-scale outflow must be due to supernovae feedback from the known starburst.

\begin{acknowledgements}

We are very grateful to the referee for very helpful and detailed comments, which improved the presentation of the paper. The ALMA staff in Chile and ARC-people at IRAM are gratefully acknowledged for their help in the data reduction. We
particularly thank Philippe Salomé for useful advice and Gerold Busch for the SINFONI data. AA acknowledges financial support by the Hellenic Foundation for Research and Innovation (HFRI) and the General Secretariat for Research and Technology (GSRT), under the project number 1882 and from the Spanish MCIU under grant ``Quantifying the impact of quasar feedback on galaxy evolution (QSOFEED)", with reference EUR2020-112266, from the Consejer{\'i}a de Econom{\'i}a, Conocimiento y Empleo del Gobierno de Canarias and the European Regional Development Fund (ERDF) under grant with reference ProID2020010105. SGB acknowledges support through grants PGC2018-094671-B-I00 (MCIU/AEI/FEDER,UE) and PID2019-106027GA-C44 from the Spanish Ministerio de Ciencia e Innovaci{\'o}n. This paper makes use of the following ALMA data: ADS/JAO.ALMA\#2015.0.00404.S, and ADS/JAO.ALMA\#2016.0.00296.S. ALMA is a partnership of ESO (representing its member states), NSF (USA) and NINS (Japan), together with NRC(Canada) and NSC and ASIAA (Taiwan), in cooperation with the Republic of Chile. The Joint ALMA Observatory is operated by ESO, AUI/NRAO and NAOJ. The National Radio Astronomy Observatory is a facility of the National Science Foundation operated under cooperative agreement by Associated Universities, Inc. We used observations made with the NASA/ESA Hubble Space Telescope, and obtained from the Hubble Legacy Archive, which is a collaboration between the Space Telescope Science Institute (STScI/NASA), the Space Telescope European Coordinating Facility (ST-ECF/ESA), and the Canadian Astronomy Data Centre (CADC/NRC/CSA). We made use of the NASA/IPAC Extragalactic Database (NED), and of the HyperLeda database. This research made use of Astropy, a community developed core Python package for Astronomy. This work was supported by the Programme National Cosmology and Galaxies(PNCG) of CNRS (INSU with INP and IN2P3), co-funded by CEA, and CNES.
\end{acknowledgements}

%-------------------------------------------------------------------
\bibliographystyle{aa} % style aa.bst
\bibliography{test.bib}

\begin{thebibliography}{77}
\expandafter\ifx\csname natexlab\endcsname\relax\def\natexlab#1{#1}\fi

\bibitem[{{Aalto} {et~al.}(1994){Aalto}, {Booth}, {Black}, {Koribalski}, \&
  {Wielebinski}}]{aalto94}
{Aalto}, S., {Booth}, R.~S., {Black}, J.~H., {Koribalski}, B., \&
  {Wielebinski}, R. 1994, \aap, 286, 365

\bibitem[{{Aalto} {et~al.}(2016){Aalto}, {Costagliola}, {Muller}, {Sakamoto},
  {Gallagher}, {Dasyra}, {Wada}, {Combes}, {Garc{\'{\i}}a-Burillo},
  {Kristensen}, {Mart{\'{\i}}n}, {van der Werf}, {Evans}, \&
  {Kotilainen}}]{aalto16}
{Aalto}, S., {Costagliola}, F., {Muller}, S., {et~al.} 2016, \aap, 590, A73

\bibitem[{{Alonso-Herrero} {et~al.}(2019){Alonso-Herrero},
  {Garc{\'\i}a-Burillo}, {Pereira-Santaella}, {Davies}, {Combes},
  {Vestergaard}, {Raimundo}, {Bunker}, {D{\'\i}az-Santos}, {Gandhi},
  {Garc{\'\i}a-Bernete}, {Hicks}, {H{\"o}nig}, {Hunt}, {Imanishi}, {Izumi},
  {Levenson}, {Maciejewski}, {Packham}, {Ramos Almeida}, {Ricci}, {Rigopoulou},
  {Roche}, {Rosario}, {Schartmann}, {Usero}, \& {Ward}}]{ah19}
{Alonso-Herrero}, A., {Garc{\'\i}a-Burillo}, S., {Pereira-Santaella}, M.,
  {et~al.} 2019, \aap, 628, A65

\bibitem[{{Alonso-Herrero} {et~al.}(2018){Alonso-Herrero}, {Pereira-Santaella},
  {Garc{\'\i}a-Burillo}, {Davies}, {Combes}, {Asmus}, {Bunker},
  {D{\'\i}az-Santos}, {Gandhi}, {Gonz{\'a}lez-Mart{\'\i}n},
  {Hern{\'a}n-Caballero}, {Hicks}, {H{\"o}nig}, {Labiano}, {Levenson},
  {Packham}, {Ramos Almeida}, {Ricci}, {Rigopoulou}, {Rosario}, {Sani}, \&
  {Ward}}]{Alonso2018}
{Alonso-Herrero}, A., {Pereira-Santaella}, M., {Garc{\'\i}a-Burillo}, S.,
  {et~al.} 2018, \apj, 859, 144

\bibitem[{{Audibert} {et~al.}(2019){Audibert}, {Combes}, {Garc{\'\i}a-Burillo},
  {Hunt}, {Eckart}, {Aalto}, {Casasola}, {Boone}, {Krips}, {Viti}, {Muller},
  {Dasyra}, {van der Werf}, \& {Mart{\'\i}n}}]{ane613}
{Audibert}, A., {Combes}, F., {Garc{\'\i}a-Burillo}, S., {et~al.} 2019, arXiv
  e-prints, arXiv:1905.01979

\bibitem[{{Awaki} \& {Koyama}(1993)}]{awaki93}
{Awaki}, H. \& {Koyama}, K. 1993, Advances in Space Research, 13, 221

\bibitem[{{Barnaby} \& {Thronson}(1992)}]{barna92}
{Barnaby}, D. \& {Thronson}, Harley~A., J. 1992, \aj, 103, 41

\bibitem[{{Bertola} {et~al.}(1991){Bertola}, {Bettoni}, {Danziger}, {Sadler},
  {Sparke}, \& {de Zeeuw}}]{ber91}
{Bertola}, F., {Bettoni}, D., {Danziger}, J., {et~al.} 1991, \apj, 373, 369

\bibitem[{{Bigiel} {et~al.}(2008){Bigiel}, {Leroy}, {Walter}, {Brinks}, {de
  Blok}, {Madore}, \& {Thornley}}]{Bigiel2008}
{Bigiel}, F., {Leroy}, A., {Walter}, F., {et~al.} 2008, \aj, 136, 2846

\bibitem[{{Bizyaev} \& {Mitronova}(2009)}]{biz09}
{Bizyaev}, D. \& {Mitronova}, S. 2009, \apj, 702, 1567

\bibitem[{{Bolatto} {et~al.}(2013){Bolatto}, {Wolfire}, \& {Leroy}}]{bol13}
{Bolatto}, A.~D., {Wolfire}, M., \& {Leroy}, A.~K. 2013, \araa, 51, 207

\bibitem[{{Busch} {et~al.}(2017){Busch}, {Eckart}, {Valencia-S.}, {Fazeli},
  {Scharw{\"a}chter}, {Combes}, \& {Garc{\'{\i}}a-Burillo}}]{bus17}
{Busch}, G., {Eckart}, A., {Valencia-S.}, M., {et~al.} 2017, \aap, 598, A55

\bibitem[{{Buta} \& {Combes}(1996)}]{buta96}
{Buta}, R. \& {Combes}, F. 1996, \fcp, 17, 95

\bibitem[{{Calzetti} {et~al.}(2000){Calzetti}, {Armus}, {Bohlin}, {Kinney},
  {Koornneef}, \& {Storchi-Bergmann}}]{calzetti00}
{Calzetti}, D., {Armus}, L., {Bohlin}, R.~C., {et~al.} 2000, \apj, 533, 682

\bibitem[{{Casasola} {et~al.}(2015){Casasola}, {Hunt}, {Combes}, \&
  {Garc{\'\i}a-Burillo}}]{vivi15}
{Casasola}, V., {Hunt}, L., {Combes}, F., \& {Garc{\'\i}a-Burillo}, S. 2015,
  \aap, 577, A135

\bibitem[{{Cicone} {et~al.}(2014){Cicone}, {Maiolino}, {Sturm},
  {Graci{\'a}-Carpio}, {Feruglio}, {Neri}, {Aalto}, {Davies}, {Fiore},
  {Fischer}, {Garc{\'{\i}}a-Burillo}, {Gonz{\'a}lez-Alfonso},
  {Hailey-Dunsheath}, {Piconcelli}, \& {Veilleux}}]{cicone14}
{Cicone}, C., {Maiolino}, R., {Sturm}, E., {et~al.} 2014, \aap, 562, A21

\bibitem[{{Collison} {et~al.}(1994){Collison}, {Saikia}, {Pedlar}, {Axon}, \&
  {Unger}}]{col94}
{Collison}, P.~M., {Saikia}, D.~J., {Pedlar}, A., {Axon}, D.~J., \& {Unger},
  S.~W. 1994, \mnras, 268, 203

\bibitem[{{Combes}(1994)}]{combes1994}
{Combes}, F. 1994, in Mass-Transfer Induced Activity in Galaxies, ed.
  I.~{Shlosman}, 170

\bibitem[{{Combes}(2001)}]{combes01}
{Combes}, F. 2001, in Advanced Lectures on the Starburst-AGN, ed.
  I.~{Aretxaga}, D.~{Kunth}, \& R.~{M{\'u}jica}, 223

\bibitem[{{Combes} {et~al.}(2019){Combes}, {Garc{\'{\i}}a-Burillo}, {Audibert},
  {Hunt}, {Eckart}, {Aalto}, {Casasola}, {Boone}, {Krips}, {Viti}, {Sakamoto},
  {Muller}, {Dasyra}, {van der Werf}, \& {Martin}}]{combes19}
{Combes}, F., {Garc{\'{\i}}a-Burillo}, S., {Audibert}, A., {et~al.} 2019, \aap,
  623, A79

\bibitem[{{Combes} {et~al.}(2014){Combes}, {Garc{\'{\i}}a-Burillo}, {Casasola},
  {Hunt}, {Krips}, {Baker}, {Boone}, {Eckart}, {Marquez}, {Neri}, {Schinnerer},
  \& {Tacconi}}]{combes1566}
{Combes}, F., {Garc{\'{\i}}a-Burillo}, S., {Casasola}, V., {et~al.} 2014, \aap,
  565, A97

\bibitem[{{Crosthwaite} \& {Turner}(2007)}]{cros07}
{Crosthwaite}, L.~P. \& {Turner}, J.~L. 2007, \aj, 134, 1827

\bibitem[{{Dahlem} {et~al.}(1990){Dahlem}, {Aalto}, {Klein}, {Booth}, {Mebold},
  {Wielebinski}, \& {Lesch}}]{Dahlem1990}
{Dahlem}, M., {Aalto}, S., {Klein}, U., {et~al.} 1990, \aap, 240, 237

\bibitem[{{Dahlem} {et~al.}(1994){Dahlem}, {Hartner}, \& {Junkes}}]{Dahlem1994}
{Dahlem}, M., {Hartner}, G.~D., \& {Junkes}, N. 1994, \apj, 432, 598

\bibitem[{{Dasyra} {et~al.}(2014){Dasyra}, {Combes}, {Novak}, {Bremer},
  {Spinoglio}, {Pereira Santaella}, {Salom{\'e}}, \& {Falgarone}}]{kal14}
{Dasyra}, K.~M., {Combes}, F., {Novak}, G.~S., {et~al.} 2014, \aap, 565, A46

\bibitem[{{de Vaucouleurs} {et~al.}(1991){de Vaucouleurs}, {de Vaucouleurs},
  {Corwin}, {Buta}, {Paturel}, \& {Fouqu{\'e}}}]{vau91}
{de Vaucouleurs}, G., {de Vaucouleurs}, A., {Corwin}, Jr., H.~G., {et~al.}
  1991, {Third Reference Catalogue of Bright Galaxies. Volume I: Explanations
  and references. Volume II: Data for galaxies between 0$^{h}$ and 12$^{h}$.
  Volume III: Data for galaxies between 12$^{h}$ and 24$^{h}$.}

\bibitem[{{Di Teodoro} \& {Fraternali}(2015)}]{barolo}
{Di Teodoro}, E.~M. \& {Fraternali}, F. 2015, \mnras, 451, 3021

\bibitem[{{Dopita} {et~al.}(2015){Dopita}, {Shastri}, {Davies}, {Kewley},
  {Hampton}, {Scharw{\"a}chter}, {Sutherland}, {Kharb}, {Jose}, {Bhatt},
  {Ramya}, {Jin}, {Banfield}, {Zaw}, {Juneau}, {James}, \&
  {Srivastava}}]{dop15}
{Dopita}, M.~A., {Shastri}, P., {Davies}, R., {et~al.} 2015, \apjs, 217, 12

\bibitem[{{Feruglio} {et~al.}(2015){Feruglio}, {Fiore}, {Carniani},
  {Piconcelli}, {Zappacosta}, {Bongiorno}, {Cicone}, {Maiolino}, {Marconi},
  {Menci}, {Puccetti}, \& {Veilleux}}]{fer15}
{Feruglio}, C., {Fiore}, F., {Carniani}, S., {et~al.} 2015, \aap, 583, A99

\bibitem[{{Feruglio} {et~al.}(2010){Feruglio}, {Maiolino}, {Piconcelli},
  {Menci}, {Aussel}, {Lamastra}, \& {Fiore}}]{fer10}
{Feruglio}, C., {Maiolino}, R., {Piconcelli}, E., {et~al.} 2010, \aap, 518,
  L155

\bibitem[{{Fluetsch} {et~al.}(2019){Fluetsch}, {Maiolino}, {Carniani},
  {Marconi}, {Cicone}, {Bourne}, {Costa}, {Fabian}, {Ishibashi}, \&
  {Venturi}}]{flu19}
{Fluetsch}, A., {Maiolino}, R., {Carniani}, S., {et~al.} 2019, \mnras, 483,
  4586

\bibitem[{{Friedli} \& {Martinet}(1993)}]{Friedli1993}
{Friedli}, D. \& {Martinet}, L. 1993, \aap, 277, 27

\bibitem[{{Garc{\'\i}a-Burillo} {et~al.}(2021){Garc{\'\i}a-Burillo},
  {Alonso-Herrero}, {Ramos Almeida}, {Gonz{\'a}lez-Mart{\'\i}n}, {Combes},
  {Usero}, {H{\"o}nig}, {Querejeta}, {Hicks}, {Hunt}, {Rosario}, {Davies},
  {Boorman}, {Bunker}, {Burtscher}, {Colina}, {D{\'\i}az-Santos}, {Gandhi},
  {Garc{\'\i}a-Bernete}, {Garc{\'\i}a-Lorenzo}, {Ichikawa}, {Imanishi},
  {Izumi}, {Labiano}, {Levenson}, {L{\'o}pez-Rodr{\'\i}guez}, {Packham},
  {Pereira-Santaella}, {Ricci}, {Rigopoulou}, {Rouan}, {Shimizu}, {Stalevski},
  {Wada}, \& {Williamson}}]{santigatos}
{Garc{\'\i}a-Burillo}, S., {Alonso-Herrero}, A., {Ramos Almeida}, C., {et~al.}
  2021, arXiv e-prints, arXiv:2104.10227

\bibitem[{{Garc{\'{\i}}a-Burillo} \& {Combes}(2012)}]{santi12}
{Garc{\'{\i}}a-Burillo}, S. \& {Combes}, F. 2012, in Journal of Physics
  Conference Series, Vol. 372, Journal of Physics Conference Series, 012050

\bibitem[{{Garc{\'\i}a-Burillo} {et~al.}(2019){Garc{\'\i}a-Burillo}, {Combes},
  {Ramos Almeida}, {Usero}, {Alonso-Herrero}, {Hunt}, {Rouan}, {Aalto},
  {Querejeta}, {Viti}, {van der Werf}, {Vives-Arias}, {Fuente}, {Colina},
  {Mart{\'\i}n-Pintado}, {Henkel}, {Mart{\'\i}n}, {Krips}, {Gratadour}, {Neri},
  \& {Tacconi}}]{santi19}
{Garc{\'\i}a-Burillo}, S., {Combes}, F., {Ramos Almeida}, C., {et~al.} 2019,
  \aap, 632, A61

\bibitem[{{Garc{\'{\i}}a-Burillo} {et~al.}(2016){Garc{\'{\i}}a-Burillo},
  {Combes}, {Ramos Almeida}, {Usero}, {Krips}, {Alonso-Herrero}, {Aalto},
  {Casasola}, {Hunt}, {Mart{\'{\i}}n}, {Viti}, {Colina}, {Costagliola},
  {Eckart}, {Fuente}, {Henkel}, {M{\'a}rquez}, {Neri}, {Schinnerer}, {Tacconi},
  \& {van der Werf}}]{santi16tor}
{Garc{\'{\i}}a-Burillo}, S., {Combes}, F., {Ramos Almeida}, C., {et~al.} 2016,
  \apjl, 823, L12

\bibitem[{{Garc{\'\i}a-Burillo} {et~al.}(2005){Garc{\'\i}a-Burillo}, {Combes},
  {Schinnerer}, {Boone}, \& {Hunt}}]{burillo2005}
{Garc{\'\i}a-Burillo}, S., {Combes}, F., {Schinnerer}, E., {Boone}, F., \&
  {Hunt}, L.~K. 2005, \aap, 441, 1011

\bibitem[{{Garc{\'{\i}}a-Burillo} {et~al.}(2014){Garc{\'{\i}}a-Burillo},
  {Combes}, {Usero}, {Aalto}, {Krips}, {Viti}, {Alonso-Herrero}, {Hunt},
  {Schinnerer}, {Baker}, {Boone}, {Casasola}, {Colina}, {Costagliola},
  {Eckart}, {Fuente}, {Henkel}, {Labiano}, {Mart{\'{\i}}n}, {M{\'a}rquez},
  {Muller}, {Planesas}, {Ramos Almeida}, {Spaans}, {Tacconi}, \& {van der
  Werf}}]{santi1068}
{Garc{\'{\i}}a-Burillo}, S., {Combes}, F., {Usero}, A., {et~al.} 2014, \aap,
  567, A125

\bibitem[{{Ginsburg} \& {Mirocha}(2011)}]{pyspec}
{Ginsburg}, A. \& {Mirocha}, J. 2011, {PySpecKit: Python Spectroscopic
  Toolkit}, Astrophysics Source Code Library

\bibitem[{{Ginsburg} {et~al.}(2015){Ginsburg}, {Robitaille}, {Beaumont},
  {Rosolowsky}, {Leroy}, {Brogan}, {Hunter}, {Teuben}, \&
  {Brisbin}}]{radiotools}
{Ginsburg}, A., {Robitaille}, T., {Beaumont}, C., {et~al.} 2015, in
  Astronomical Society of the Pacific Conference Series, Vol. 499, Revolution
  in Astronomy with ALMA: The Third Year, ed. D.~{Iono}, K.~{Tatematsu},
  A.~{Wootten}, \& L.~{Testi}, 363--364

\bibitem[{{Guilloteau} \& {Lucas}(2000)}]{gildas}
{Guilloteau}, S. \& {Lucas}, R. 2000, in Astronomical Society of the Pacific
  Conference Series, Vol. 217, Imaging at Radio through Submillimeter
  Wavelengths, ed. J.~G. {Mangum} \& S.~J.~E. {Radford}, 299

\bibitem[{{Ho} {et~al.}(2011){Ho}, {Li}, {Barth}, {Seigar}, \& {Peng}}]{ho11}
{Ho}, L.~C., {Li}, Z.-Y., {Barth}, A.~J., {Seigar}, M.~S., \& {Peng}, C.~Y.
  2011, \apjs, 197, 21

\bibitem[{{Hopkins} \& {Quataert}(2010)}]{hop10}
{Hopkins}, P.~F. \& {Quataert}, E. 2010, \mnras, 407, 1529

\bibitem[{{Hunt} {et~al.}(2008){Hunt}, {Combes}, {Garc{\'\i}a-Burillo},
  {Schinnerer}, {Krips}, {Baker}, {Boone}, {Eckart}, {L{\'e}on}, {Neri}, \&
  {Tacconi}}]{hunt2008}
{Hunt}, L.~K., {Combes}, F., {Garc{\'\i}a-Burillo}, S., {et~al.} 2008, \aap,
  482, 133

\bibitem[{{Izumi} {et~al.}(2016){Izumi}, {Kohno}, {Aalto}, {Espada}, {Fathi},
  {Harada}, {Hatsukade}, {Hsieh}, {Imanishi}, {Krips}, {Mart{\'{\i}}n},
  {Matsushita}, {Meier}, {Nakai}, {Nakanishi}, {Schinnerer}, {Sheth},
  {Terashima}, \& {Turner}}]{izumi16}
{Izumi}, T., {Kohno}, K., {Aalto}, S., {et~al.} 2016, \apj, 818, 42

\bibitem[{{Jim{\'e}nez-Bail{\'o}n} {et~al.}(2005){Jim{\'e}nez-Bail{\'o}n},
  {Santos-Lle{\'o}}, {Dahlem}, {Ehle}, {Mas-Hesse}, {Guainazzi}, {Heckman}, \&
  {Weaver}}]{Jimenez-Bailon2005}
{Jim{\'e}nez-Bail{\'o}n}, E., {Santos-Lle{\'o}}, M., {Dahlem}, M., {et~al.}
  2005, \aap, 442, 861

\bibitem[{{Kennicutt} {et~al.}(2007){Kennicutt}, {Calzetti}, {Walter}, {Helou},
  {Hollenbach}, {Armus}, {Bendo}, {Dale}, {Draine}, {Engelbracht}, {Gordon},
  {Prescott}, {Regan}, {Thornley}, {Bot}, {Brinks}, {de Blok}, {de Mello},
  {Meyer}, {Moustakas}, {Murphy}, {Sheth}, \& {Smith}}]{kenni07}
{Kennicutt}, Robert~C., J., {Calzetti}, D., {Walter}, F., {et~al.} 2007, \apj,
  671, 333

\bibitem[{{Kennicutt}(1998)}]{kenni98}
{Kennicutt}, Jr., R.~C. 1998, \araa, 36, 189

\bibitem[{{Koribalski} {et~al.}(1993){Koribalski}, {Dahlem}, {Mebold}, \&
  {Brinks}}]{kor93}
{Koribalski}, B., {Dahlem}, M., {Mebold}, U., \& {Brinks}, E. 1993, \aap, 268,
  14

\bibitem[{{Koribalski} {et~al.}(1996){Koribalski}, {Dettmar}, {Mebold}, \&
  {Wielebinski}}]{kor96}
{Koribalski}, B., {Dettmar}, R.-J., {Mebold}, U., \& {Wielebinski}, R. 1996,
  \aap, 315, 71

\bibitem[{{Kotilainen} {et~al.}(1996){Kotilainen}, {Forbes}, {Moorwood}, {van
  der Werf}, \& {Ward}}]{koti96}
{Kotilainen}, J.~K., {Forbes}, D.~A., {Moorwood}, A.~F.~M., {van der Werf},
  P.~P., \& {Ward}, M.~J. 1996, \aap, 313, 771

\bibitem[{{Kruijssen} {et~al.}(2018){Kruijssen}, {Schruba}, {Hygate}, {Hu},
  {Haydon}, \& {Longmore}}]{kruijssen2018}
{Kruijssen}, J.~M.~D., {Schruba}, A., {Hygate}, A. e. P.~S., {et~al.} 2018,
  \mnras, 479, 1866

\bibitem[{{McMullin} {et~al.}(2007){McMullin}, {Waters}, {Schiebel}, {Young},
  \& {Golap}}]{casa}
{McMullin}, J.~P., {Waters}, B., {Schiebel}, D., {Young}, W., \& {Golap}, K.
  2007, in Astronomical Society of the Pacific Conference Series, Vol. 376,
  Astronomical Data Analysis Software and Systems XVI, ed. R.~A. {Shaw},
  F.~{Hill}, \& D.~J. {Bell}, 127

\bibitem[{{O'Neill} \& {Dubinski}(2003)}]{oneill2003}
{O'Neill}, J.~K. \& {Dubinski}, J. 2003, \mnras, 346, 251

\bibitem[{{Panuzzo} {et~al.}(2003){Panuzzo}, {Bressan}, {Granato}, {Silva}, \&
  {Danese}}]{pan03}
{Panuzzo}, P., {Bressan}, A., {Granato}, G.~L., {Silva}, L., \& {Danese}, L.
  2003, \aap, 409, 99

\bibitem[{{Phillips}(1993)}]{phi93}
{Phillips}, A.~C. 1993, \aj, 105, 486

\bibitem[{{Privon} {et~al.}(2015){Privon}, {Herrero-Illana}, {Evans},
  {Iwasawa}, {Perez-Torres}, {Armus}, {D{\'\i}az-Santos}, {Murphy},
  {Stierwalt}, {Aalto}, {Mazzarella}, {Barcos-Mu{\~n}oz}, {Borish}, {Inami},
  {Kim}, {Treister}, {Surace}, {Lord}, {Conway}, {Frayer}, \&
  {Alberdi}}]{privon15}
{Privon}, G.~C., {Herrero-Illana}, R., {Evans}, A.~S., {et~al.} 2015, \apj,
  814, 39

\bibitem[{{Reif} {et~al.}(1982){Reif}, {Mebold}, {Goss}, {van Woerden}, \&
  {Siegman}}]{reif82}
{Reif}, K., {Mebold}, U., {Goss}, W.~M., {van Woerden}, H., \& {Siegman}, B.
  1982, \aaps, 50, 451

\bibitem[{{Robitaille} \& {Bressert}(2012)}]{aplpy}
{Robitaille}, T. \& {Bressert}, E. 2012, {APLpy: Astronomical Plotting Library
  in Python}, Astrophysics Source Code Library

\bibitem[{{Rogstad} {et~al.}(1974){Rogstad}, {Lockhart}, \& {Wright}}]{rog74}
{Rogstad}, D.~H., {Lockhart}, I.~A., \& {Wright}, M.~C.~H. 1974, \apj, 193, 309

\bibitem[{{Rupke} {et~al.}(2005){Rupke}, {Veilleux}, \& {Sanders}}]{rupke05}
{Rupke}, D.~S., {Veilleux}, S., \& {Sanders}, D.~B. 2005, \apj, 632, 751

\bibitem[{{Saikia} {et~al.}(1990){Saikia}, {Unger}, {Pedlar}, {Yates}, {Axon},
  {Wolstencroft}, {Taylor}, \& {Gyldenkerne}}]{saikia90}
{Saikia}, D.~J., {Unger}, S.~W., {Pedlar}, A., {et~al.} 1990, \mnras, 245, 397

\bibitem[{{Salak} {et~al.}(2016){Salak}, {Nakai}, {Hatakeyama}, \&
  {Miyamoto}}]{salak16}
{Salak}, D., {Nakai}, N., {Hatakeyama}, T., \& {Miyamoto}, Y. 2016, \apj, 823,
  68

\bibitem[{{Salak} {et~al.}(2019){Salak}, {Nakai}, {Seta}, \&
  {Miyamoto}}]{salak19}
{Salak}, D., {Nakai}, N., {Seta}, M., \& {Miyamoto}, Y. 2019, \apj, 887, 143

\bibitem[{{Salak} {et~al.}(2017){Salak}, {Tomiyasu}, {Nakai}, {Kuno},
  {Miyamoto}, \& {Kaneko}}]{salak17}
{Salak}, D., {Tomiyasu}, Y., {Nakai}, N., {et~al.} 2017, \apj, 849, 90

\bibitem[{{Salak} {et~al.}(2018){Salak}, {Tomiyasu}, {Nakai}, {Kuno},
  {Miyamoto}, \& {Kaneko}}]{salak2018}
{Salak}, D., {Tomiyasu}, Y., {Nakai}, N., {et~al.} 2018, \apj, 856, 97

\bibitem[{{Sanders} {et~al.}(2003){Sanders}, {Mazzarella}, {Kim}, {Surace}, \&
  {Soifer}}]{san03}
{Sanders}, D.~B., {Mazzarella}, J.~M., {Kim}, D.-C., {Surace}, J.~A., \&
  {Soifer}, B.~T. 2003, \aj, 126, 1607

\bibitem[{{Schruba} {et~al.}(2010){Schruba}, {Leroy}, {Walter}, {Sand strom},
  \& {Rosolowsky}}]{schruba2010}
{Schruba}, A., {Leroy}, A.~K., {Walter}, F., {Sand strom}, K., \& {Rosolowsky},
  E. 2010, \apj, 722, 1699

\bibitem[{{Schuster} {et~al.}(2007){Schuster}, {Kramer}, {Hitschfeld},
  {Garcia-Burillo}, \& {Mookerjea}}]{schuster07}
{Schuster}, K.~F., {Kramer}, C., {Hitschfeld}, M., {Garcia-Burillo}, S., \&
  {Mookerjea}, B. 2007, \aap, 461, 143

\bibitem[{{S{\'e}rsic} \& {Pastoriza}(1965)}]{sp65}
{S{\'e}rsic}, J.~L. \& {Pastoriza}, M. 1965, \pasp, 77, 287

\bibitem[{{Shlosman} {et~al.}(1989){Shlosman}, {Frank}, \&
  {Begelman}}]{Shlosman1989}
{Shlosman}, I., {Frank}, J., \& {Begelman}, M.~C. 1989, \nat, 338, 45

\bibitem[{{Solomon} \& {Vanden Bout}(2005)}]{sol05}
{Solomon}, P.~M. \& {Vanden Bout}, P.~A. 2005, \araa, 43, 677

\bibitem[{{Steer} {et~al.}(2017){Steer}, {Madore}, {Mazzarella}, {Schmitz},
  {Corwin}, {Chan}, {Ebert}, {Helou}, {Baker}, {Chen}, {Frayer}, {Jacobson},
  {Lo}, {Ogle}, {Pevunova}, \& {Terek}}]{steer17}
{Steer}, I., {Madore}, B.~F., {Mazzarella}, J.~M., {et~al.} 2017, \aj, 153, 37

\bibitem[{{Veilleux} {et~al.}(2020){Veilleux}, {Maiolino}, {Bolatto}, \&
  {Aalto}}]{Veilleux2020}
{Veilleux}, S., {Maiolino}, R., {Bolatto}, A.~D., \& {Aalto}, S. 2020, \aapr,
  28, 2

\bibitem[{{Veilleux} {et~al.}(2013){Veilleux}, {Mel{\'e}ndez}, {Sturm},
  {Gracia-Carpio}, {Fischer}, {Gonz{\'a}lez-Alfonso}, {Contursi}, {Lutz},
  {Poglitsch}, {Davies}, {Genzel}, {Tacconi}, {de Jong}, {Sternberg}, {Netzer},
  {Hailey-Dunsheath}, {Verma}, {Rupke}, {Maiolino}, {Teng}, \&
  {Polisensky}}]{veilleux13}
{Veilleux}, S., {Mel{\'e}ndez}, M., {Sturm}, E., {et~al.} 2013, \apj, 776, 27

\bibitem[{{Veron-Cetty} \& {Veron}(1986)}]{veron86}
{Veron-Cetty}, M.-P. \& {Veron}, P. 1986, \aaps, 66, 335

\bibitem[{{Wong} \& {Blitz}(2002)}]{wong02}
{Wong}, T. \& {Blitz}, L. 2002, \apj, 569, 157

\end{thebibliography}

% The total CO(3-2) integrated spectrum is shown in Fig. \ref{tot}, and the total CO flux is computed, with the help of 3 gaussian fits. 

%\section{Integrated spectrum and total mass estimation}

% \begin{figure}
%  \resizebox{\hsize}{!}{\includegraphics{co_tot+fit.png}}
% \caption[Integrated CO(3-2) spectrum.]{We show the total CO(3-2) emission line profile integrated over the observed map, with a FOV of 18\arcsec\, after correction for primary beam attenuation. The light blue line is the result of the Gaussian fit with three velocity components (in dark blue), to better compute the total flux.}
% \label{tot}
% \end{figure}

%\begin{table}
%\caption{Line fluxes}              % title of Table
%\label{flux}      
%\centering                                      
%\begin{tabular}{c c c c c}         
%\hline\hline                       
%Line & $S_{CO}$ & V & $\Delta$V$^a$  & ${S_{peak}}^b$ \\  
%        & (Jy.km/s)  &   (km/s) &  (km/s) & (Jy) \\
%\hline                                  
%C1 & 658.1 $\pm$ 113.8  &  -111.3 $\pm$ 1.7  &  78.1 $\pm$ 5.3  &  7.9 %\\
%C2 & 1673.3 $\pm$ 187.6  &  0.5 $\pm$ 4.2  &  164.2 $\pm$ 15.8  &  9.6 %\\
%C3 & 263.0 $\pm$ 53.7  &  132.5 $\pm$ 2.0  &  60.6 $\pm$ 6.2  &  4.1 %\\
%\hline        
%%HCN(4-3) & \\
%%HCO(4-3) & \\
%%CS(7-6) & \\
%\end{tabular}
%\\
%\tablefoot{
%Results of the Gaussian fits for the total CO(3-2) emission assuming %the 3 velocity components (C1, C2 and C3), shown in Fig.~\ref{tot}. \\
%\tablefoottext{$\rm ^a$}{Full width at half maximum (FWHM)} \\
%\tablefoottext{$\rm ^b$}{Peak flux}
%}
%\end{table}

\begin{appendix}

\section{Channel maps}
\begin{figure*}[b]
\centering
\includegraphics[width=17cm]{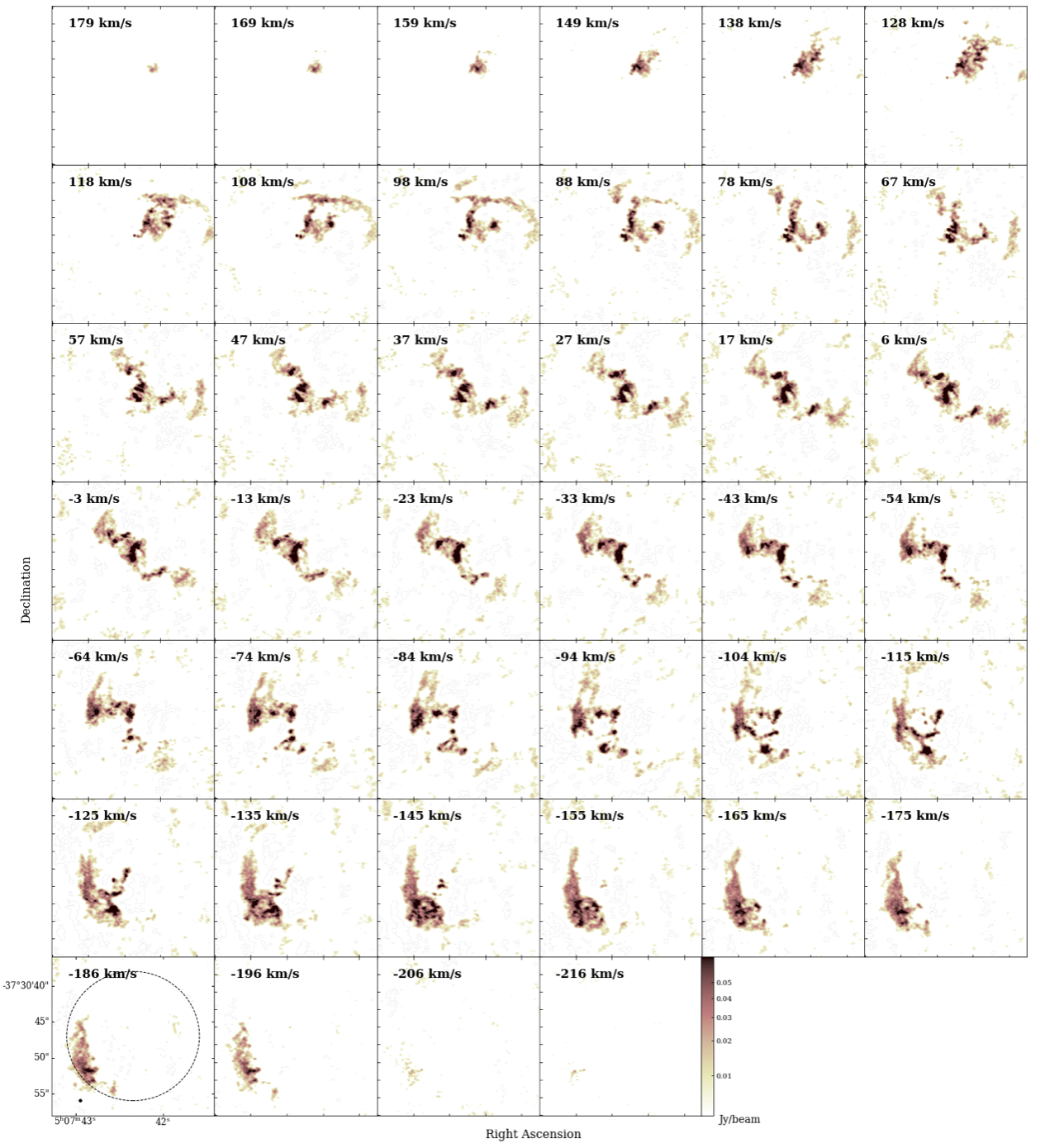}
\caption{Channel maps of CO(3-2) emission in the centre of NGC\,1808. Each of the 40 square boxes is 22\arcsec\ in size, while the FWHP=16.9\arcsec\ of the primary beam primary beam is indicated by the dotted circle in the bottom left panel. Channels are separated by 10.2km/s. They are displayed from 179 (top left) to -216\,km/s (bottom right) related the $v(sys)=995$\,km$\rm s^{-1}$. The synthesized beam size is 0. 29\arcsec$\times$0. 24\arcsec\ (PA=79.6$^\circ$) and it is displayed as the black ellipse at the bottom left panel. The centre of the maps is the phase centre of the interferometric observations given in Table~\ref{prop}. The color scale is in power stretch, ranging between 0 and 70\,mJy/beam.}
\label{chans}
\end{figure*}

In Figure~\ref{chans}, we present the channel maps of the CO(3-2) emission.

\end{appendix}

\end{document}